\documentclass[apj]{emulateapj}
\shorttitle{SN~2011dh}
\newcommand{\Ni}{{$^{56}$Ni}}

\shortauthors{Bersten et al.}
\begin{document}
\title{The Type IIb Supernova 2011dh from a Supergiant Progenitor} 
\author{Melina C. Bersten\altaffilmark{1}, Omar
  G. Benvenuto\altaffilmark{2}\footnote{OGB is member of
    the Carrera del 
Investigador Cient\'{\i}fico de la Comisi\'on de Investigaciones
Cient\'{\i}ficas 
de la Provincia de Buenos Aires (CIC), Argentina.}, Ken'ichi
  Nomoto\altaffilmark{1},  
Mattias Ergon\altaffilmark{3}, Gast\'on Folatelli\altaffilmark{1},
Jesper Sollerman\altaffilmark{3}, Stefano Benetti\altaffilmark{5},
Maria Teresa Botticella\altaffilmark{4}, Morgan
Fraser\altaffilmark{6}, Rubina Kotak\altaffilmark{6}, Keiichi
Maeda\altaffilmark{1},  Paolo Ochner\altaffilmark{5}, Lina
Tomasella\altaffilmark{5}}    

\affil{\altaffilmark{1} Kavli Institute for the Physics and Mathematics of
  the Universe, Todai Institutes for Advanced Study, University of
  Tokyo, 5-1-5 Kashiwanoha, Kashiwa, Chiba 277-8583, Japan} 
\affil{\altaffilmark{2} Facultad de Ciencias Astron\'omicas y
  Geof\'{\i}sicas, Universidad Nacional de La Plata, Paseo del Bosque
  S/N, B1900FWA La Plata, Argentina}  

\affil{\altaffilmark{3} The Oskar Klein Centre, Department of
  Astronomy, AlbaNova, SE-106 91 Stockholm, Sweden}

\affil{\altaffilmark{4}INAF-Osservatorio Astronomico di Capodimonte
Salita Moiariello, 16  80131 - Napoli}  

\affil{\altaffilmark{5}INAF-Osservatorio Astronomico di Padova, vicolo
  dell'Osservatorio 5, 35122 Padova, Italy}

\affil{\altaffilmark{6}Astrophysics Research Centre, School of
  Mathematics and Physics, Queen's University Belfast, Belfast BT7 1NN}

\email{melina.bersten@ipmu.jp}
\submitted{Submitted to ApJ on May 24, 2012 --- Accepted on July 17,
2012.}

\begin{abstract}
\noindent
A set of hydrodynamical models based on
stellar evolutionary progenitors is used to study the nature
of SN~2011dh. Our modeling suggests that a large progenitor star
---with $R \sim 200$ $R_\odot$---, is needed to reproduce the 
early light curve of SN~2011dh. This is consistent with
  the suggestion that the yellow super-giant star detected at the
  location of the SN in deep pre-explosion images is the progenitor star.
From the main peak of the bolometric light curve and expansion
velocities we constrain the mass of the ejecta to be $\approx$ $2$
$M_\odot$, the explosion energy to be  $E= 6-10\times 10^{50}$ erg, and
the \Ni\, mass  to be approximately $0.06$ $M_\odot$.  The progenitor star was
composed of a helium core of  3 to 4 $M_\odot$ and a thin
hydrogen-rich envelope of $\approx$ $0.1$
  $M_\odot$ with a main sequence mass estimated to be 
 in the range of 12--15 $M_\odot$. Our
models rule out progenitors with helium-core masses larger than 8
$M_\odot$, which correspond to $M_{\mathrm{ZAMS}} \gtrsim 25$
$M_\odot$. This suggests that a single star evolutionary scenario for
SN~2011dh is unlikely. 
\end{abstract}

\keywords{hydrodynamics---supernovae: general --- supernovae:
  individual: SN~2011dh }

\section{INTRODUCTION}
\label{sec:intro}
Type IIb supernovae (SNe~IIb) are transitional objects within the
family of core-collapse SNe (CCSNe), as their spectroscopic classification
evolves from Type II (i.e. with H lines), to type Ib (i.e. dominated
by helium lines). SNe~IIb were recognized as a new class with the
discoveries of SN~1987K \citep{1988AJ.....96.1941F} and SN~1993J 
\citep[e.g.][]{1993Natur.364..507N,1997ARA&A..35..309F}. As the 
 rest of CCSNe---which includes H-rich types II-P and II-L, H-less
 type Ib, and He-less type Ic---, they are believed to arise from 
the violent death of stars with initial masses greater than 8
$M_\odot$. Massive stars may suffer considerable mass loss during  
their evolution, due to strong stellar winds or mass transfer to a binary
companion. Therefore, the vast spectroscopic and photometric diversity
observed among CCSNe is related to the ability of the progenitor
star to retain its  outermost layers before the
explosion. Despite the efforts
  to improve  our understanding of the progenitor of each subclass of
CCSNe, many  questions remain open. In this
sense, the study of a very well-observed object can shed light on the
nature of CCSNe and  their massive progenitor systems.

SN~2011dh was discovered on May 31.893 by amateur astronomers and
immediately  
confirmed by the Palomar Transient Factory (PTF) in the nearby spiral
galaxy M51
\citep{2011CBET.2736....1G,2011ATel.3398....1S,2011ApJ...742L..18A}
which had also hosted three other CCSNe in the past 17 years.  
Strong constraints on the date of explosion, to better than $0.6$ days were
established using pre- and post-SN imaging
\citep{2011ApJ...742L..18A}. SN~2011dh
was extensively monitored in a wide wavelength range, including early
detections in radio and X-rays \citep{2012ApJ...752...78S}. It
was classified as type IIb \citep{2011ApJ...742L..18A}, a relatively
rare subclass of CCSNe, based on the optical
  spectrum. Soon after discovery, a source was identified 
as a possible progenitor of  SN~2011dh in archival, multi-band HST images
\citep{2011ApJ...739L..37M,2011ApJ...741L..28V}. Photometry of the
source is compatible with a yellow super-giant (YSG) star.
 
The detection of pre-supernova (pre-SN) objects in high-resolution imaging has
provided important information on the possible progenitors of several
SNe. For SNe~II-P, these observations confirm the red supergiant
nature of the progenitor, as  previously suggested by  theory. An
upper limit of  $M_\mathrm{ZAMS}= 16.5 \pm 1.5 \, M_\odot$ was
suggested for this type of SNe \citep{2009MNRAS.395.1409S}. 
However, such upper limit has been recently revised by
\citet{2012MNRAS.419.2054W}. They found a higher value for the maximum
mass of SN~II-P progenitors of $M_\mathrm{ZAMS}= 21^{+2}_{-1}\,
M_\odot$ when additional extinction due to dust produced in the red
supergiant wind is taken into account. For other subtypes of CCSNe,
detections have not been as common and only in a few cases
have the progenitors been conclusively identified, e.g the type II-pec 
SN~1987A \citep{1987Natur.328..318G}, the type IIb SN~1993J 
\citep{1994AJ....107..662A,2004Natur.427..129M}, and the type IIn SN~2005gl
\citep{2007ApJ...656..372G}. 

To determine the main-sequence mass ($M_\mathrm{ZAMS}$) from 
pre-SN imaging, it is necessary to derive a luminosity and intrinsic
colour from the photometry, and compare with some
evolutionary track. This was done for SN~2011dh by
\citet{2011ApJ...739L..37M} and \citet{2011ApJ...741L..28V}. But although
both studies obtained consistent effective temperature and 
luminosity, and they employed the same evolutionary
model, the derived $M_\mathrm{ZAMS}$ was different. Because color
uncertainties are expected to arise from unknown mass-loss history of
the progenitor, \citet{2011ApJ...739L..37M} assumed only the
luminosity was reliable, and derived 
$M_\mathrm{ZAMS}= 13\pm 3 \; M_\odot$ from the end point of the
evolutionary track that matched the luminosity.  
Meanwhile  \citet{2011ApJ...741L..28V} derived $M_\mathrm{ZAMS}=
17-19 \; M_\odot$ by choosing the closest track that matches
the luminosity and color of the source in the Hertzsprung-Russell
(H-R) diagram although this point does not correspond with the final
position of the star at the end of its evolution.
Alternatively, \citet{2011ApJ...742L...4M} compared
stellar population synthesis 
with the stellar association surrounding SN~2011dh to derive
$M_\mathrm{ZAMS}$. Assuming that the stars in the vicinity of the SN
are coeval, they concluded that  the value of $M_\mathrm{ZAMS}$ is
most likely close to the estimate of \citet{2011ApJ...739L..37M}. 

Some authors have suggested that the YSG star detected in the pre-SN images
is not the actual progenitor of SN~2011dh but its binary companion or
even an unrelated object. These authors suggest,
instead, that the exploding object was a compact star. The arguments
for this are based, first, 
on the shock velocity derived from radio and 
sub-millimeter observations 
\citep{2012ApJ...752...78S,2012arXiv1201.0771B,2012arXiv1201.0770K}. The
relatively high shock velocity found, $v_{\mathrm{sw}} \approx 0.1\,
c$, would indicate a compact progenitor, or a
type cIIb SN, in the scheme proposed by
\citet{2010ApJ...711L..40C}. The second argument for a compact
progenitor was introduced by \citet{2011ApJ...742L..18A} and is based
on the quick decline of the optical light curve (LC) soon after the
explosion, as compared with SN~1993J, along with a low temperature
derived from an early-time spectrum, as compared with analytic
expressions by \citet{2011ApJ...728...63R}. Such a compact progenitor
would be inconsistent with the YSG star identified by
\citet{2011ApJ...739L..37M} and \citet{2011ApJ...741L..28V}, which is
expected to have a radius of $\approx 270 \,R_\odot$ (from
$\mathrm{log}\, L= 4.92$ $L_\odot$,
$T_{\mathrm{eff}}= 6000 \,K$, given $L= R^2\,
T_{\mathrm{eff}}^4$ in solar units). 

 Early observations such as those available for SN~2011dh provide a
  unique opportunity to analyze the physical properties of the
  progenitor. In particular, one of the most direct ways to
  estimate the size of the progenitor is by modeling the LC during the
  cooling phase that occurs after the shock
  break-out and before the re-heating by radioactive
  decay. Observations during this phase are very scarce due to its short 
  duration and so they are very valuable. The studies described above do
  not present a specific modeling of the early light curve. With the
  aim of assessing the nature of the progenitor we set out to perform
  a more detailed modeling of the available observations.

A well-known method to estimate the properties of the progenitor
object, as well as the explosion energy, and the amount and  
distribution of the radioactive material, is to compare the observations
with predictions from  hydrodynamical models. In this paper,  
we present a set of hydrodynamical models applied to stellar
evolutionary progenitors that aim to elucidate 
the compact or extended nature of the progenitor of SN~2011dh, as well
as the main physical parameters of the explosion. Our initial and LC
models  
are presented in \S~\ref{sec:models}. A comparison between models and
observations is done in \S~\ref{sec:LC}. The global physical
parameters are studied in \S~\ref{sec:SLC}, and the sensitivity of the
LCs on the initial radius is investigated in \S~\ref{sec:ELC}. In 
 \S~\ref{sec:discussion} we discuss the results and summarize our main
 conclusions \S~\ref{sec:conclusion}.

\section{Stellar and Supernova models}
\label{sec:models}

To study the nature of SN~2011dh we compare the observations with a
set of hydrodynamical models applied to initial structures from stellar
evolutionary calculations. Our supernova models are 
computed using a one-dimensional Lagrangian hydrodynamic code with
flux-limited radiation diffusion \citep{2011ApJ...729...61B}. The
explosion itself is simulated  by injecting a certain amount
of energy near the center of the progenitor object, which produces a
powerful shock wave that propagates through the 
progenitor transforming thermal and kinetic energy of the matter into
energy that can be radiated from the stellar surface. During the
propagation of the shock wave, explosive nucleosynthesis
produces unstable isotopes of iron-group elements, mainly \Ni. The 
decay of \Ni\,$\longrightarrow$$^{56}$Co$\longrightarrow$$^{56}$Fe
produces energetic $\gamma$-rays and positrons that are thermalized,
 providing extra energy that contributes to power the LC. The code
 includes $\gamma$-ray transfer in gray approximation for any
 distribution of \Ni\,, assuming a constant value for the
 $\gamma$-ray opacity, $\kappa_\gamma=0.06\, y_e$ cm$^2$g$^{-1}$, where
 $y_e$ is the number of electrons per baryon
 \citep{1995ApJ...446..766S}. Note that this approximation 
  is appropriate for typical SN conditions as it
   has been shown by \citet{1995ApJ...446..766S} through comparison
   with Monte Carlo simulations.
Nucleosynthesis is, however, not  consistently calculated but it is 
included as a pre-explosion condition, assuming that the energy 
released does not contribute significantly to
the dynamics of the explosion itself, and that it only effect 
is on the chemical structure  \citep[see][for
further details of this assumption]{WW90,1996snih.book.....A}. 

LCs of SNe~IIb have been successfully reproduced using helium
 stars with a very thin hydrogen envelope ($\lesssim 1 \, M_\odot$)
as pre-SN models
\citep{1994ApJ...420..341S,1994ApJ...429..300W,1998ApJ...496..454B}.
 Therefore,  
we adopt the same type of model as input for our hydrodynamic  
calculations. The He core models were calculated by \citet{NH88}
using a single-star evolutionary code that  follows 
the evolution of the He core from He burning until the collapse of the
core assuming solar initial abundance. The external H-rich
envelope of the He core was replaced with the proper boundary
conditions obtained from the hydrostatic and thermal equilibrium
envelope models \citep{NS72,N74}.
To take into account the thin hydrogen
envelope, we smoothly attached a low-mass H-rich envelope
in hydrostatic and thermal equilibrium to the He core. The presence of
the external envelope significantly modifies the radius but
not the mass of the progenitor. 

Specifically, we employ four different initial models with He core
masses of 3.3 $M_\odot$ (He3.3), 4 $M_\odot$ (He4), 5 $M_\odot$
(He5), and 8 $M_\odot$ (He8), which correspond to the stellar
evolution of single stars with main-sequence masses of 12, 15, 18, and
25 $M_\odot$, respectively \citep[see][for the He core mass --
  main-sequence mass relation]{SN80,2009ApJ...692.1131T}. 
   Figure~\ref{fig:rho} shows the initial density profile as a function
  of radius for the He4 model and for models with different H-rich
  envelopes attached to the core of the He4 model. The latter models will be
  analyzed in \S~\ref{sec:ELC}.
  
The chemical
structure of these He stars before the explosion consists 
of a Fe core of $\approx \,$1.4 $M_\odot$, surrounded by concentric
layers of  
 Si-rich, O-rich, and He-rich material. For each of the initial models, 
explosive nucleosynthesis was calculated with a reaction network that
includes 280 isotopes up to $^{79}$Br
\citep{1996ApJ...460..869H,1999ApJ...511..862H}. The results of the
nucleosynthesis depend on the progenitor mass and the kinetic 
energy of the explosion and they have already been presented in
previous studies \citep{1996ApJ...460..408T}. An important
  characteristic of the models is the variation in the mass of the
O-rich layer which goes from $0.2$  $M_\odot$ for the He3.3 model to 
 $3\, M_\odot$ for the He8 model. Therefore, the determination of the
oxygen mass using late-time spectra may provide a 
constraint on the main-sequence mass of the progenitor
\citep{2010Natur.465..326K}.  

In Section~\ref{sec:SLC} we study the explosion models for the three
least massive initial models (He3.3, He4, and He5), using several values
of explosion energy, \Ni\, mass, and \Ni\, distribution. In this analysis
the mixing of \Ni\, is modified from that of the nucleosynthesis
calculations in order to explain the rising part of the observed LC,
as frequently done to model the LC of SNe~Ibc 
\citep[e.g. see][]{1994ApJ...420..341S,1994ApJ...429..300W,2000ApJ...532.1132B}. 
Such a mixing between the core and the He layer in Type Ib and IIb 
supernovae is caused by the Rayleigh-Taylor instability at the base of
the He layer \citep{H91,H94,I97}. Actually, more recent 3D
  numerical calculations predict even larger mixing of \Ni\, than
  previous 2D calculations
  \citep{2010ApJ...714.1371H,2010ApJ...723..353J}. This result is in
  better agreement with the extensive mixing observed in
  e.g. SN~1987A.  
In Section~\ref{sec:ELC} we study the effect of the external H-rich
envelope which must be present to explain the H spectral features
observed at early times. Finally, the He8 model is discussed in
Section~\ref{sec:discussion} to test the plausibility of a massive single
progenitor star.

\section{Light Curve Modeling}
\label{sec:LC}
The $g'$-band LC of SN~2011dh showed a quick decline during 1--3 days
after an initial peak with $M_g^{\mathrm{peak}}= -16.5 $ mag. This
initial decline was 
followed by a re-brightening in all bands leading to a second peak
with bolometric luminosity of $L_{\mathrm{bol}}\approx 1.4 \times
10^{42}$ erg s$^{-1}$ ($M_{\mathrm{bol}}= -16.6$ mag) at $t \approx 20$
days. Double-peaked LC have been   
observed for very few SNe, one of which was the well-known SN~1993J,
although with a slower initial decline and larger luminosity, as
compared with SN~2011dh. The scarcity of observations during the early
declining phase is due to its very short duration,
which makes it difficult to catch.  
Theoretically, however,
all CCSNe should present this early behavior which is a consequence of
cooling after the shock  breaks out from the surface of the object. 
 The duration of the cooling phase is mainly regulated by the size of
 the progenitor,  
 although there is also 
 a dependence on energy, mass and envelope composition. More compact
 structures, such as Wolf-Rayet stars, produce faster declines than
 extended progenitors, such as red supergiants.

On the other hand, the heat source that powers the second peak is
provided by the radioactive decay of \Ni\, produced during the
explosive nucleosynthesis and its daughter $^{56}$Co. Modeling the LC
around the second peak provides information about the explosion energy
($E$), ejecta mass ($M_{\mathrm{ej}}$), and the mass and distribution
of \Ni\,. Such global properties for SN~2011dh are analyzed in
Section~\ref{sec:SLC} while the effect of the progenitor radius which
essentially affects the early decline phase of the LC is studied in
Section~\ref{sec:ELC}. 

The observed bolometric LC of SN~2011dh and the expansion velocities for
the Fe~II lines used in this paper are taken from
\citet{E12}. A distance of 7.1 Mpc \citep{2006MNRAS.372.1735T}, a
Galactic reddening of $E(B-V)= 0.035$ mag and no host-galaxy reddening
\citep{2011ApJ...742L..18A} were assumed in the calculations.  
For the early cooling phase, there are only a few observations available
in a limited wavelength range, which prevents
us from calculating a bolometric luminosity during that epoch. Hence
we directly adopt the $g'$-band data published by
\citet{2011ApJ...742L..18A}. In the present work we assume the 
explosion time was 2011 May $31.5$ UT, that is the mid-point between
the last non-detection and the discovery date.

\subsection{Global properties of SN~2011dh}
\label{sec:SLC} 
To  estimate physical parameters, such as explosion energy ($E$),
ejected mass ($M_{\mathrm{ej}}$), and the mass (\Ni\,mass) and
distribution of \Ni, we analyze the bolometric LC around the second
peak, and the photospheric velocitiy ($v_{\mathrm{ph}}$)
evolution. For this purpose, the He star models presented in
section~\ref{sec:models} are adopted as initial configurations. To
save computation time, no external H envelope was added at this point
because this does not affect the LC around the main peak. Therefore,
the progenitor is assumed to be compact with a radius of $R \approx
R_\odot$. 

First we analyze the general dependence of the LC and
$v_{\mathrm{ph}}$ on variations of the physical parameters for model
He4. This is the He core used by \citet{1994ApJ...420..341S} to
model the bolometric LC of SN~1993J with $E \approx 1.2$ foe (1 foe=
$1 \times 10^{51}$erg), \Ni\, mass of $\sim$$0.08$ $M_\odot$, and
assuming substantial mixing of \Ni\, out to the He layers (see also
our own model of SN~1993J in Figure~\ref{fig:LC93J},
Section~\ref{sec:compact}). Following
that work, for SN~2011dh we test physical parameters near those values. 

Figures~\ref{fig:LCVE} and \ref{fig:VE}
show the effect of the explosion energy on the LC and on
$v_{\mathrm{ph}}$, respectively, for 
energies of $E= 0.8$, $1.0$, and $1.5$ foe. More energetic explosions
produce larger kinetic and radiative energy, which is reflected in
higher photospheric velocity and global brightness. After the cooling
phase, all LCs evolve through a minimum with luminosity
$L_{\mathrm{min}}$, followed by a broad second peak with maximum luminosity
$L_{\mathrm{peak}}$, and then they enter a more or less linear
decline, or ``tail'' phase. In the following analysis we denote the
epochs of $L_{\mathrm{min}}$ 
and $L_{\mathrm{peak}}$ as $t_{\mathrm{min}}$ and $t_{\mathrm{peak}}$,
respectively. From Figure~\ref{fig:LCVE} we see that
$L_{\mathrm{min}}$ and $L_{\mathrm{peak}}$ increase with $E$ while the
luminosity during the tail phase shows a  dependence in the opposite
direction. Note that $t_{\mathrm{min}}$ is more or less independent of
explosion energy, but $t_{\mathrm{peak}}$ decreases with $E$. Finally,
models with lower $E$ produce broader second peaks.

The effect of the \Ni\, mass on the LC is shown in
Figure~\ref{fig:LCVNiM} for \Ni\, mass $= 0.05$, $0.06$, and $0.07$
$M_\odot$. The LCs are remarkably equal until about 10 days before
  $t_{\mathrm{peak}}$ when the models begin to differ in the sense
that larger \Ni\, mass produces larger luminosity and  wider
LC. $L_{\mathrm{peak}}$ and  the  tail luminosity 
increase with \Ni\, mass but no appreciable effect is seen in
$L_{\mathrm{min}}$. Also note that $t_{\mathrm{min}}$ and
$t_{\mathrm{peak}}$ are essentially insensitive to the amount of
synthesized \Ni.

The distribution of \Ni\, could be different than the one predicted by
one-dimensional nucleosynthesis calculations where it is concentrated in the inner 
regions of the progenitor. Mixing of \Ni\, out to the helium envelope
is expected to occur because of Rayleigh-Taylor instabilities during the
propagation of the shock wave which are not properly taken into
account in our one-dimensional prescription. To study the effect of mixing, we  
calculated three models with \Ni\, linearly mixed (in mass
  coordinate) out to $75\%$ (He4Mix75), 
$85\%$ (He4Mix85), and $95\%$ (He4Mix95) of the total initial mass. 
 Note that the distributions of \Ni\, adopted are in good concordance
  with recent 3D calculations \citep[see e.g. Fig.~6
    of][]{2010ApJ...714.1371H}. 

The resulting LCs are shown in
Figure~\ref{fig:LCVNiD}. The main effect of mixing is seen at $t <
t_{\mathrm{peak}}$ although there is also a smaller effect on the tail
luminosity. For more extensive mixing, $t_{\mathrm{min}}$ is produced
at earlier times  because of the earlier heating by 
radioactive decay. This also produces a higher $L_{\mathrm{min}}$,
while the value of $L_{\mathrm{peak}}$ seems not to be very sensitive
to mixing. Instead the effect is more visible
  for the LC width which  increases with more extended mixing. 

The results of these tests show that the effect on $v_{\mathrm{ph}}$
of the \Ni\, mass and distribution is essentially
negligible. Therefore, the kinetic energy available in the ejecta is
the only relevant parameter that determines the photospheric velocity
evolution, which imposes strict constraints on $E$ for a given
$M_{\mathrm{ej}}$. 

This comparative analysis allows us to find a set of parameters that
reproduce the observations of SN~2011dh. For He4, these
parameters are $E=1$ foe, and \Ni\,mass$\,= 0.065$  $M_\odot$, mixed out
to $95\%$ of the total mass or out to 9000 km s$^{-1}$ in velocity
space. Note that although \Ni\,
was mixed far out in the ejecta, since the adopted mixing function was
linear, the amount of \Ni\, in the outer layers was
small ($< 1.3 \times 10^{-3}$ $M_\odot$ for $v  > 5000$ km
s$^{-1}$). We adopted a cut mass 
($M_{\mathrm{cut}}$) of $1.5$ $M_\odot$ that is assumed to form a
compact remnant. Therefore, the mass of the ejecta is
$M_{\mathrm{ej}}= M_{\mathrm{total}}-M_{\mathrm{cut}}= 2.5 M_\odot$. 
A summary of the parameters used in this
and the following calculations is presented in Table~\ref{tbl-1}, where
we show the main-sequence mass, 
He core mass, and radius of the progenitor, the mass cut  assumed, the
ejecta mass, the explosion energy and the $^{56}\mathrm{Ni}$ mass.

We also tested the He3.3 and He5 models with smaller and larger pre-SN
mass, respectively. The parameters found for these cases are
slightly different from those of the He4 model (see Table~\ref{tbl-1}),
although the \Ni\, mixing was kept up to $95\%$ of the total mass.
For instance, in order to keep the agreement with the observed
velocities, we modified the explosion energy so that the ratio
$E/M_{\mathrm{ej}}$ remains approximately constant.
Figures~\ref{fig:LC1} and \ref{fig:velo} show the LCs
and $v_{\mathrm{ph}}$, respectively, for the  
parameters of models He3.3, He4, and He5 compared with the
observations. From the figures, it is clear that the three models give
good fits to the observations,  with the two least massive ones
  producing slightly better results. 
 
 Models He3.3 and He4 give the best fits to velocity and LC,
  respectively. This means that there is a slight tension in the
  solution of the progenitor properties based on the comparison of our
  calculations with these observables. However, given the
  uncertainties involved, it is likely that the 
  best global fit arises from a model with parameters in the range of
  these two models. That is, $M_{\mathrm{ej}}= 1.8$--$2.5$ $M_\odot$,
$E=0.6$--$1.0$ foe, and \Ni\,mass$\,= 0.060$ --$0.065$ $M_\odot$ mixed out to a
velocity of $\approx$ 9000 km s$^{-1}$. Such a progenitor with a
 He core mass between 3 and 4 $M_\odot$ corresponds to a star of
$\approx$ 12--15 $M_\odot$ on the main sequence. Although there are
 uncertanties in the determination of the main sequence mass from the
 He core mass, depending for example on convective overshooting,
 rotation, etc., for the range of He core masses considered here
 different stellar calculations provide similar results (see Figure~1
 of Tanaka et al. 2009). Stars of such initial masses
have no known mechanism to almost completely remove the H envelope, as
required for producing a SN~IIb, except through binary interaction. In
Section~\ref{sec:discussion} we study the possibility of more massive
progenitors and discuss the binary scenario.

\subsection{Progenitor radius}
\label{sec:ELC}
\begin{figure}
\begin{center}
\includegraphics[scale=.60,angle=-90]{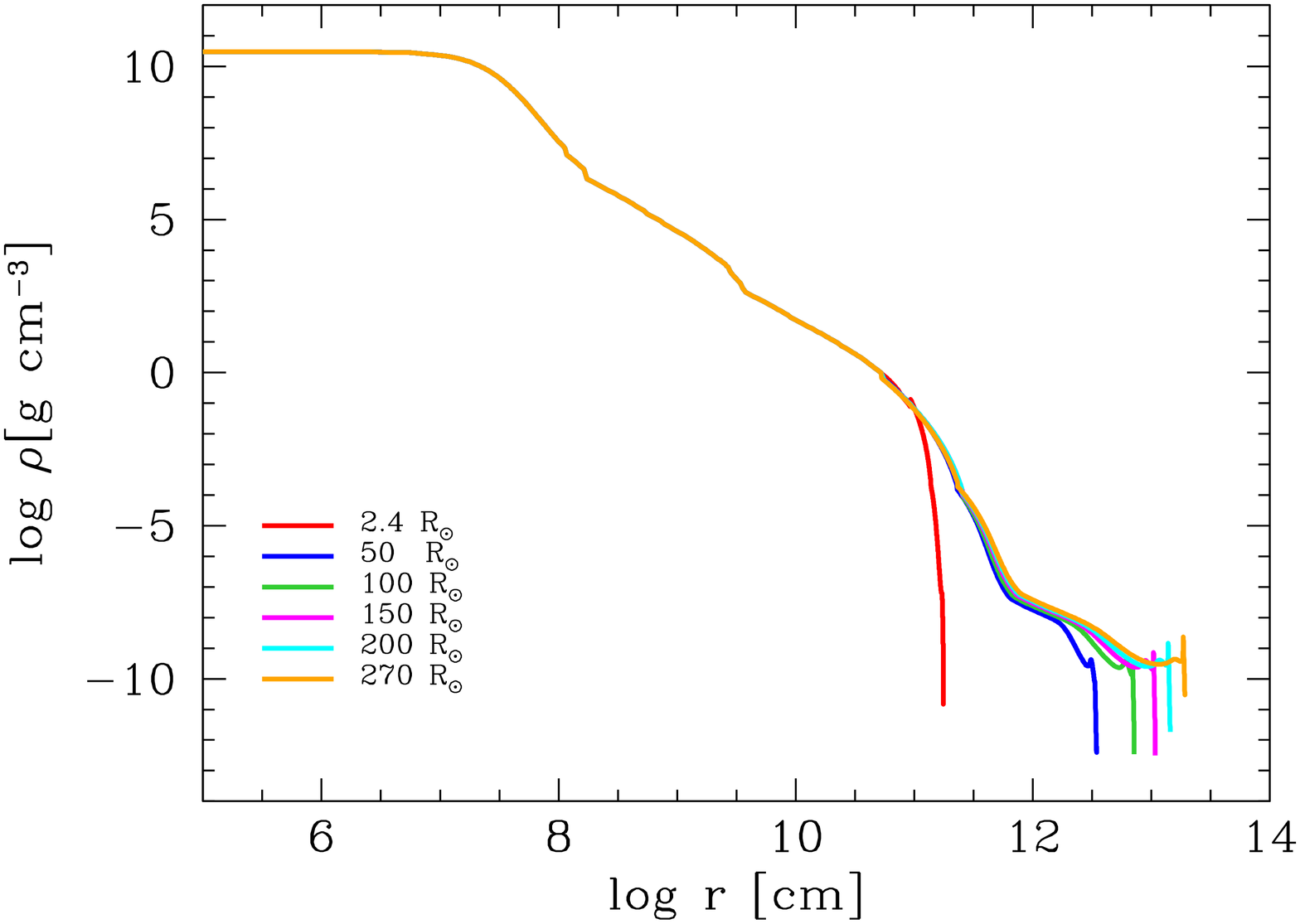}
\caption{The initial density distributions as a function of
  radius for model He4 (red line) and for models with different H-rich
  envelopes attached to the core of He4. \label{fig:rho}}
\end{center}
\end{figure}

\begin{figure}
\begin{center}
\includegraphics[scale=.60,angle=-90]{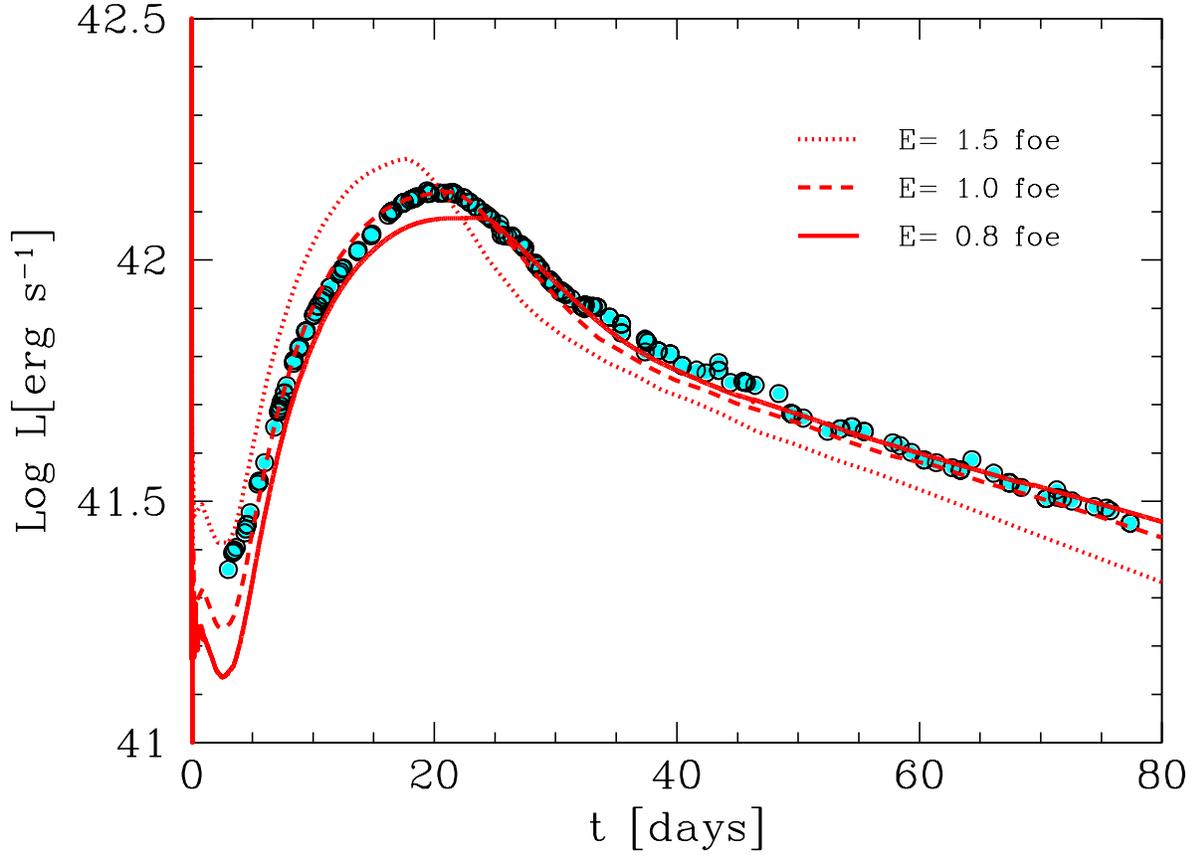}
\caption{Sensitivity of the bolometric LC to changes in the explosion
  energy. The He4 initial model (see \S~\ref{sec:SLC}) and three different
  values of the explosion energy, $E= 0.8, 1.0, 1.5$ foe, were used in these
  simulations. The observed bolometric LC of SN~2011dh (points) is
  shown for comparison.
\label{fig:LCVE}}
\end{center}
\end{figure}

\begin{figure}
\begin{center}
\includegraphics[scale=.60,angle=-90]{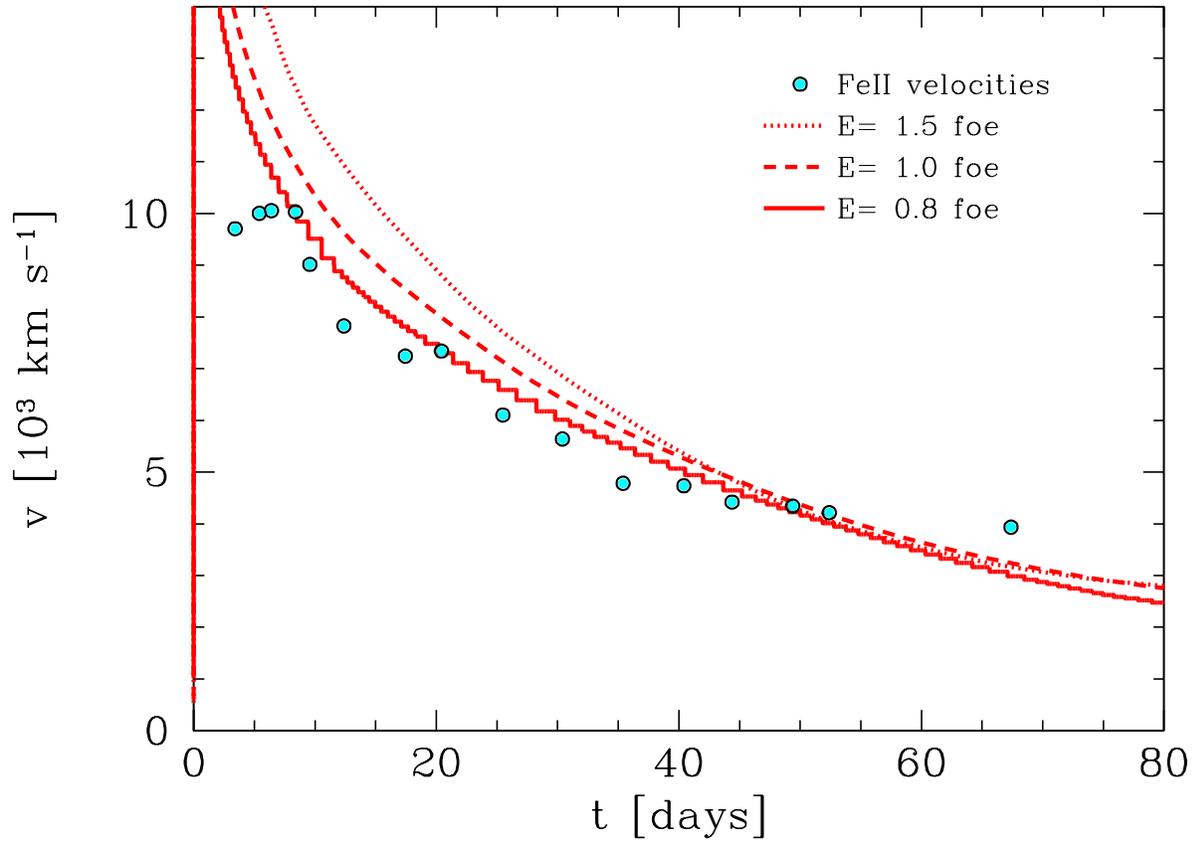}
\caption{Sensitivity of the photospheric velocity evolution on the
  explosion energy. The He4 initial model (see \S~\ref{sec:SLC}) and three
  different values of the explosion energy, $E= 0.8, 1, 1.5$ foe, were
  used in these simulations. Fe~II line velocities measured from spectra
  of SN~2011dh (points) are shown for comparison.  
\label{fig:VE}}
\end{center}
\end{figure}

\begin{figure}
\begin{center}
\includegraphics[scale=.60,angle=-90]{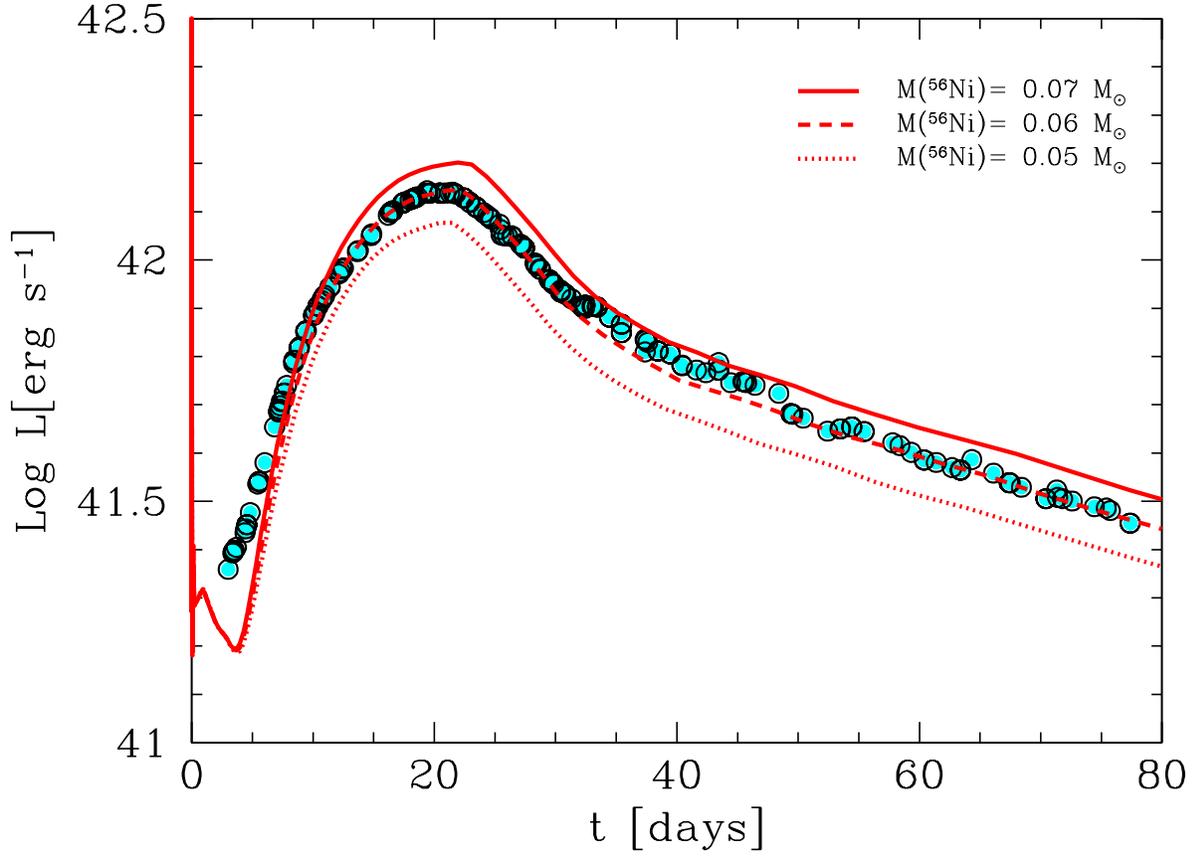}
\caption{Sensitivity of the bolometric LC to changes in \Ni\, mass.  The
  He4 initial model (see \S~\ref{sec:SLC}) and three different values of the
  \Ni\, mass, \Ni\,mass $= 0.05, 0.06, 0.07$ $M_\odot$, were used in these
  simulations. The observed bolometric LC of SN~2011dh (points) is
  shown for comparison.  
\label{fig:LCVNiM}}
\end{center}
\end{figure}

\begin{figure}
\begin{center}
\includegraphics[scale=.60,angle=-90]{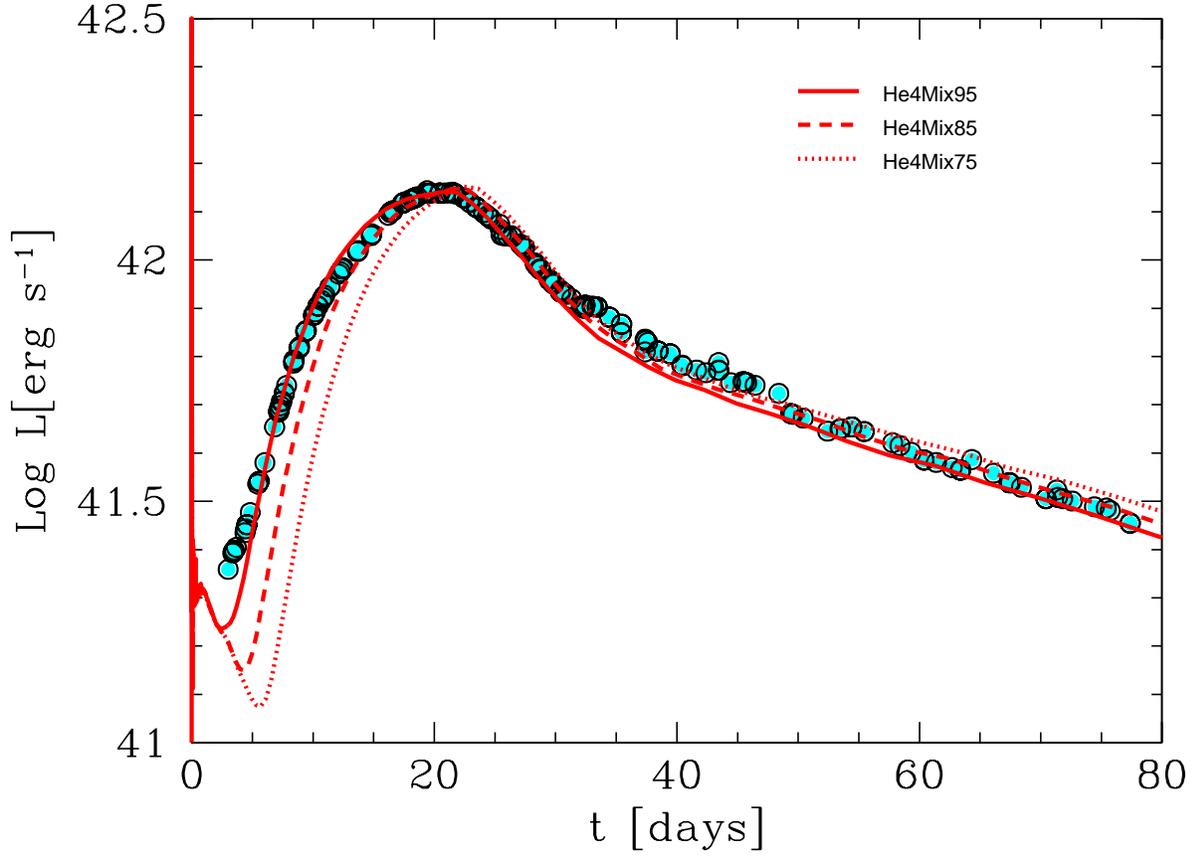}
\caption{Sensitivity of the bolometric LC to changes in the \Ni\,
  distribution. The He4 initial model with \Ni\,mass= 0.06 $M_\odot$
  (see \S~\ref{sec:SLC}) and three 
  different degrees of mixing, up to $75\%$ (He4Mix75), 
$85\%$ (He4Mix85), and $95\%$ (He4Mix95) of the total initial mass,
  were used in these simulations. The observed bolometric LC of
  SN~2011dh (points) is shown for comparison. 
\label{fig:LCVNiD}}
\end{center}
\end{figure}

\begin{figure}
\begin{center}
\includegraphics[scale=.60,angle=-90]{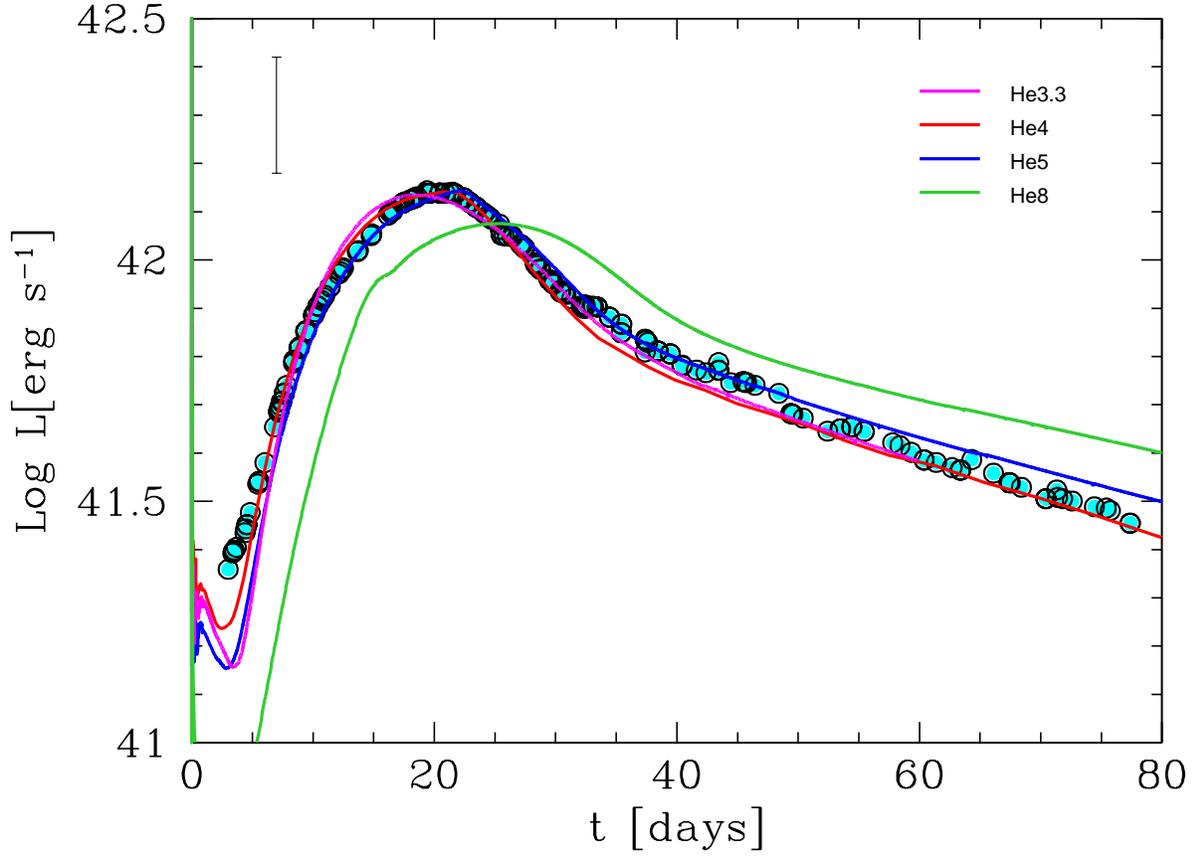}
\caption{Observed bolometric LC of SN~2011dh (points) compared with
  the results of the 
  LC calculations for models He3.3 (magenta solid line), He4 (red solid
  line) and He5 (blue solid line). An extra model,  
 He8 (green solid line) is also included to show that larger helium
 core mass is not compatible with the observations. The error bar
 indicates the size of the systematic uncertainty in luminosity as
 derived from an uncertainty of 1 Mpc in the distance to M51.
The  physical parameters used in each simulation are given in
Table~\ref{tbl-1}. 
\label{fig:LC1}}
\end{center}
\end{figure}

\begin{figure}
\begin{center}
\includegraphics[scale=.60,angle=-90]{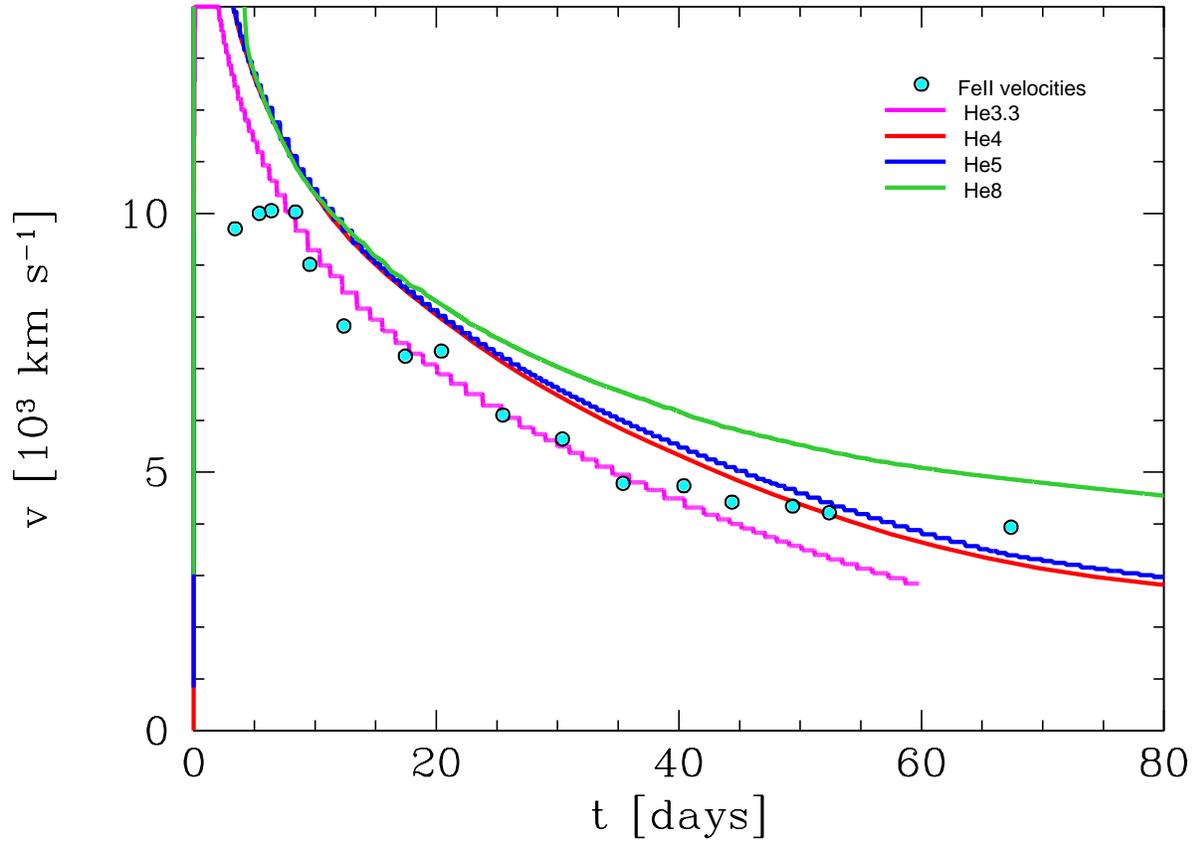}
\caption{Evolution of the photospheric velocity for models He3.3
  (magenta solid line), He4 (red solid line), He5 (blue solid line),
and He8 (green solid line) compared with measured Fe~II line
velocities of SN~2011dh. The physical parameters used in each
simulation are given in Table~1. Note that model with larger  
helium mass overestimate the observed velocities.  
\label{fig:velo}}
\end{center}
\end{figure}

To test the effect of the progenitor radius on the LC, it is critical
to obtain observations during the adiabatic cooling phase before the
radioactive decay becomes the main source to power the
LC. For the case of SN~2011dh, only a few data points in the $g'$-band are
available at those times. Here we compare these data with two
hydrodynamical models: a compact 
one with an initial radius of $\approx$ 2 $R_\odot$, and an extended one
with $R \approx 270$ $R_\odot$, consistent with the $L$ and $T_{\mathrm{eff}}$
of the YSG star detected in pre-SN images.  Both models 
have a He core mass of 4 $M_\odot$ and the same physical parameters as
model He4 (see Table~\ref{tbl-1}). The  extended model (He4R270) was
constructed by smoothly attaching to the helium core a H/He envelope
in hydrostatic and thermal equilibrium (see Figure~\ref{fig:rho}). The 
mass ($M_\mathrm{env}$) and the helium mass fraction ($Y$) of the
envelope for a uniform composition were fit so that the
 $L$ and $T_{\mathrm{eff}}$  became equal to the estimate from
pre-SN imaging. This yielded values of $M_\mathrm{env}= 0.1$ $M_\odot$
and $Y=0.8$. Note that the total H mass in this model is
$\approx$ $0.02$ $M_\odot$, similar to the value derived from spectral
modeling \citep{2011ApJ...742L..18A}.

\begin{figure}
\begin{center}
\includegraphics[scale=.40]{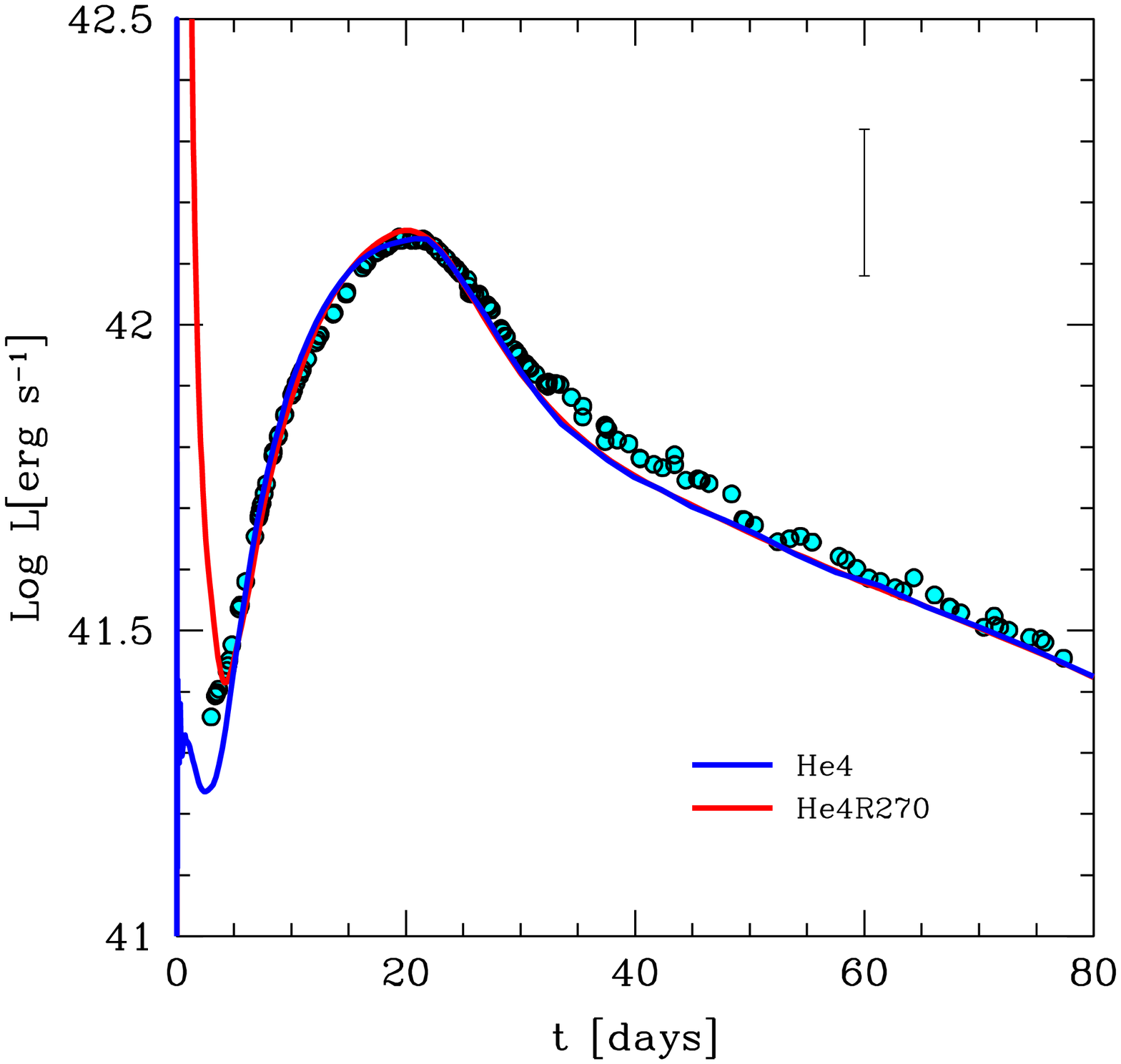}\includegraphics[scale=.40]{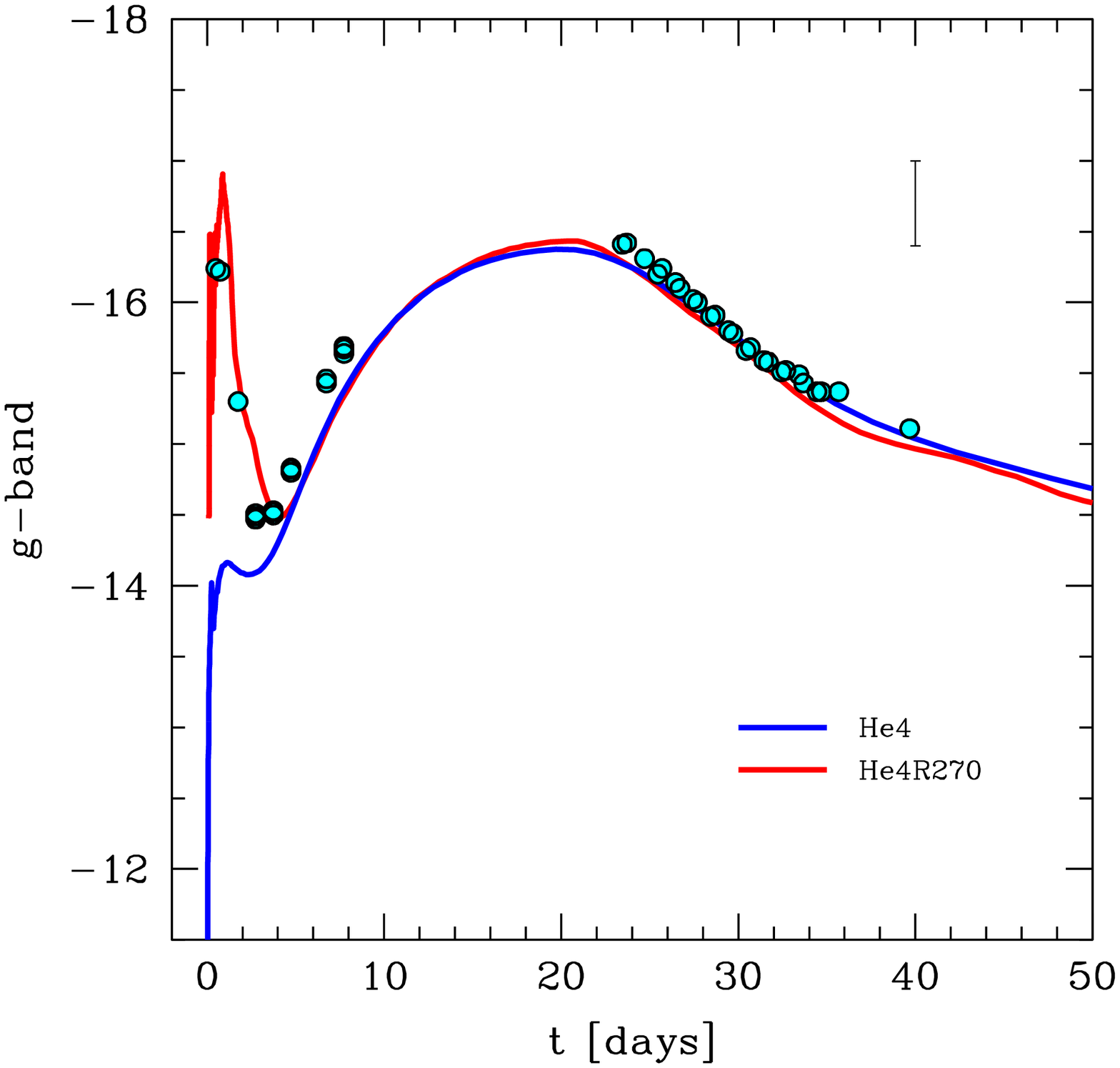}
\caption{Observed and modeled bolometric LCs {\bf(left panel)} and
 absolute  $g'$-band LCs {\bf(right panel)}. The dots show the observed
  bolometric LC from \citet{E12}, and the $g'$-band LC
  from \citet{2011ApJ...742L..18A}. The blue solid lines show the
  results for the compact progenitor model He4. The red solid lines
  corresponds to an extended progenitor, He4R270, consistent with a
  YSG star as detected in pre-SN photometry of SN~2011dh. 
Clearly, the effect of the progenitor radius is only important
before the radioactive material becomes the main
source of radiative power. The larger radius is necessary to reproduce
the early part of the $g'$-band LC. The error bars
 indicate the size of the systematic uncertainty that corresponds to 
 an uncertainty of 1 Mpc in the distance to M51.\label{fig:LCR2R270}}  
\end{center}
\end{figure}

Figure~\ref{fig:LCR2R270} (left panel) shows the bolometric LCs for
both the compact and 
the extended progenitor models. The effect
of radius is only noticeable before $t \approx 5$ days, which is
  roughly where the bolometric LC data begin. At $t\sim1$ day the
  difference in bolometric luminosity between the two models is very
  large, of up to one order of magnitude. This difference is mostly due
  to the extra amount of energy required to expand a more compact
    structure. Fortunately, earlier 
data obtained in the $g'$ band are available to discriminate between
the two models. Therefore, to compare with the $g'$-band data we calculated
synthetic LCs assuming black-body emission and integrating the flux
through the corresponding filter transmission function.
 Figure~\ref{fig:LCR2R270} (right panel) shows a comparison of our models 
with the $g'$-band observations for
compact (blue line) and extended (red line) progenitors. Again, both
  models differ significantly in the $g'$ band at $t<5$ days. While the
  extended model produces a pronounced spike, in agreement with the
  observations, the compact progenitor shows a much weaker bump. From
  this analysis we conclude that an extended progenitor is favored by
  the early-time observations.

\begin{figure}
\begin{center}
\includegraphics[scale=.60,angle=-90]{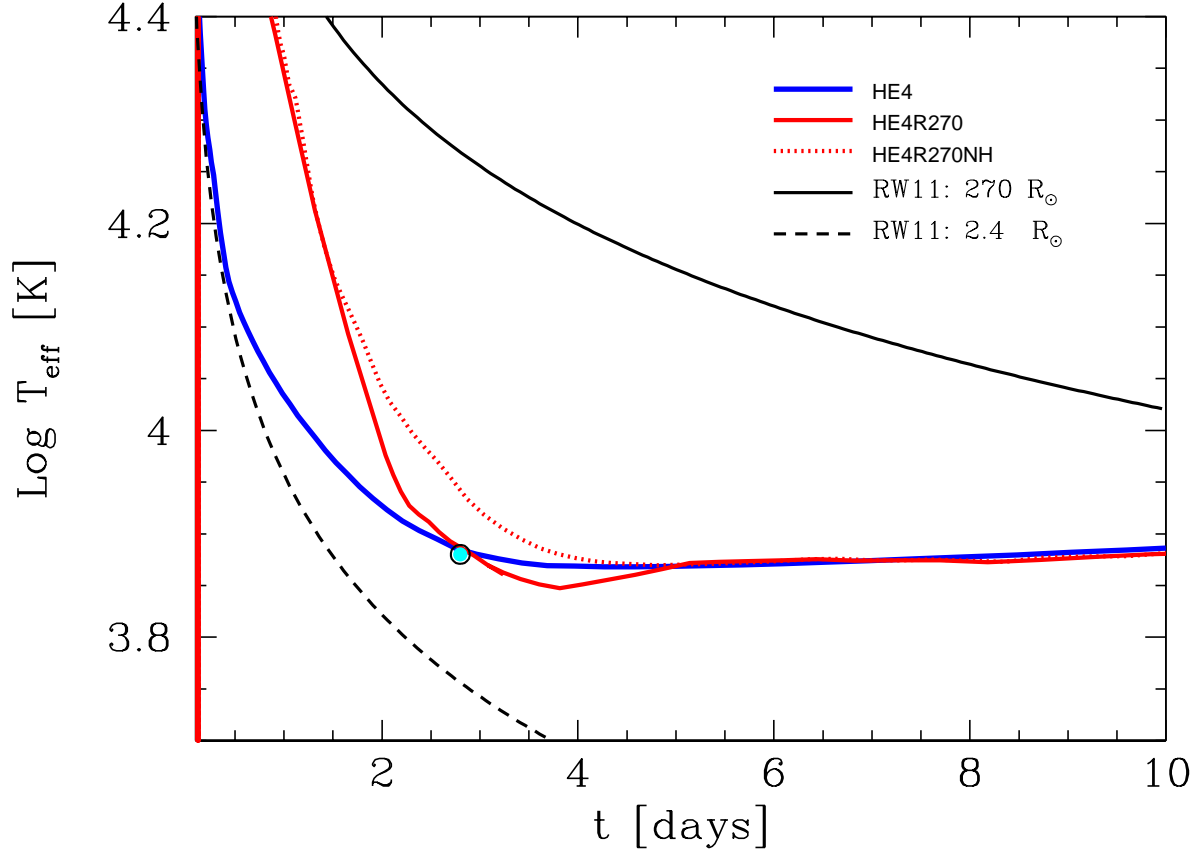}
\caption{Time evolution of the effective temperature for the compact
  model He4 and the extended models He4R270 and He4R270NH (see
  \S~\ref{sec:ELC}). The effective temperature calculated using the
  analytic expression of \citet{2011ApJ...728...63R} with $R= 270
  \, R_\odot$  (solid black line) and $R= 2.4 \, R_\odot$  (dashed black
  line) are also shown. The black-body temperature (cyan dot)
  estimated from a spectrum of SN~2011dh obtained at $2.8$ days is
  included for comparison.  
 \label{fig:Teff}}
\end{center}
\end{figure}

Further evidence for this scenario can be sought by analyzing the
temperature of the ejecta. Figure~\ref{fig:Teff} shows the evolution
of effective temperature for both models.  There are no
important differences in the effective temperature for $t \gtrsim 2$
days, contrary to what was suggested by
\citet{2011ApJ...742L..18A}. In that paper, the authors estimated a
black-body temperature of $\approx$ 7600 K from the spectrum of
SN~2011dh at $2.8$ days, and 
compared it with the effective temperature derived from an analytic
expression of 
\citep[][RW11 hereafter]{2011ApJ...728...63R} that strongly depends on
the progenitor 
radius, and less strongly on other parameters. Assuming a radius of
$\sim$$10^{13}$ cm at $t= 2.8$ 
days after the explosion, the predicted temperature is $T_{\mathrm{eff}}
\approx 17000$ K. This discrepancy is one of the  main arguments they
used to rule out an extended 
progenitor, such as a YSG. On the contrary, our numerical
calculations give values of $T_{\mathrm{eff}}=$ 7730 K and
7690 K at this epoch for 
compact and extended progenitor, respectively, which is compatible
with the value derived from the spectrum. 

An additional model named
He4R270NH is shown in Figure~\ref{fig:Teff} (dotted line). This model
is similar to He4R270, but with larger helium mass fraction in
  the envelope of $Y \approx 1$, i.e. essentially hydrogen-free. 
 One can see that the presence of H in
  the envelope of the extended progenitor (model He4R270) improves the
  agreement with the observation by reducing the temperature, as compared with a
  model where no H is included (model He4R270NH). The 
  composition of the envelope thus plays an important role in the
  determination of the 
  effective temperature. 

For comparison, Figure~\ref{fig:Teff} also shows the behavior
of $T_\mathrm{eff}$ calculated using equation (13) of RW11 for the
cases of compact and extended progenitors. The adopted radii are the
same as those employed for the hydrodynamical calculations, i.e. $R =
2.4 \, R_\odot$ and $R = 270 \, R_\odot$. Note that only at very early
times do the analytic models agree with the hydrodynamical
calculations. At later times the agreement 
breaks down. This could be partially due to the effect of
recombination in the ejecta which is not considered in expression (13)
of RW11. Another possible reason for the disagreement could be
  differences in the initial density structures between both
  formulations (see Figure~\ref{fig:rho} for our initial density
  profile). A more 
  detailed study is needed to clarify the origin of the discrepancy,
  but that goes beyond the scope of the present work.
In any case, it is important to bear in mind that the analytic expressions
were applied at an epoch ($\approx$ $2.8$ days) that could be outside
their validity range.

In any case, the $T_{\mathrm{eff}}$ is not directly comparable with
the black-body temperature derived from the spectrum. A more direct
comparison can be done using the 
color temperature ($T_{\mathrm{C}}$). Following the prescription
of \citet{1992ApJ...393..742E} we estimated $T_{\mathrm{C}}$ as the
temperature at the ``thermalization'' depth\footnote{The
  ``thermalization'' depth 
is calculated as the layer where $3 \,\tau_{\mathrm{abs}}
\, \tau_{\mathrm{sct}} \approx 1$, where $\tau_{\mathrm{sct}}$ is the
optical depth for scattering and $\tau_{\mathrm{abs}}$ is the optical
depth for absorption.
}, which led to values of 8500 K and 8300 K at $2.8$ days for models He4
and He4R270, respectively. Although these values are somewhat higher
than the value estimated from the spectrum, the discrepancy is not
important given the uncertainties in
the time of explosion ($\sim$ $0.6$ day) and in the estimations of the
color temperatures. Because of the small differences in temperature found at
$t\sim 2$ days between compact and
expended progenitors, the available temperature measurement is not a
suitable discriminator between these scenarios. 

\begin{figure}
\begin{center}
\includegraphics[scale=.40]{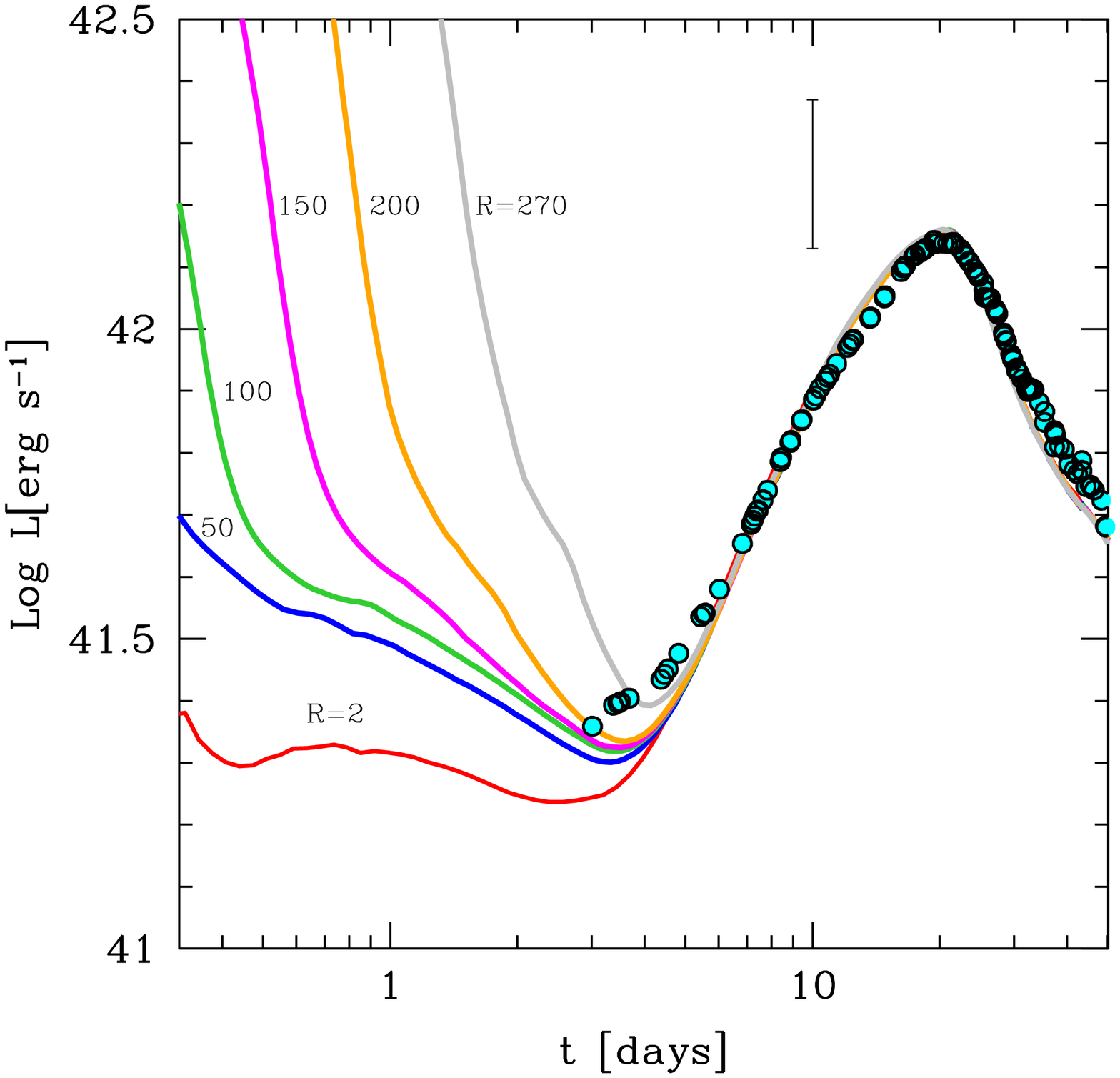}\includegraphics[scale=.40]{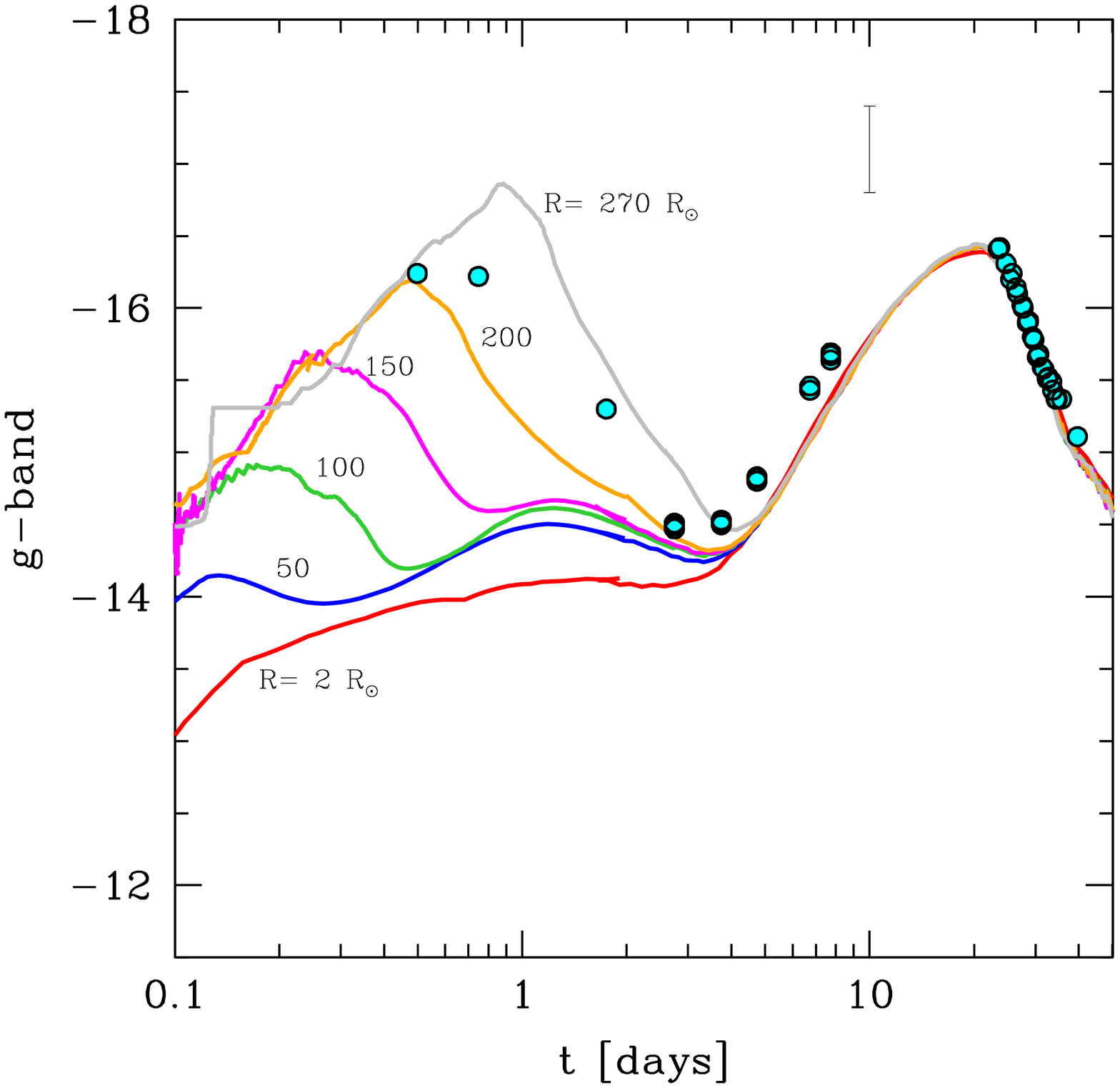}
\caption{Bolometric LCs {\bf(left panel)} and $g'$-band LCs  
{\bf(right panel)} for models with the same explosion energy as
  our preferred model, but different initial radii. The observed
  bolometric LC \citep{E12} and $g'$-band LC
  \citep{2011ApJ...742L..18A} of SN~2011dh (cyan dots) are shown for
  comparison in each panel. The error bars indicate the size of the
  systematic uncertainty that corresponds to an uncertainty of 1 Mpc
  in the distance to M51. The radius variation is accomplished by
  attaching essentially massless ($<0.01$ $M_\odot$) envelopes to the
  He4 model. Larger radii produce higher 
  early luminosity for $t \lesssim 5$ days but no
  appreciable effect is seen at later times. \label{fig:LCVR}}
\end{center}
\end{figure}

Finally, we analyzed whether it is possible to improve the comparison
between models and early observations assuming different values of the
progenitor radius than that inferred for the YSG
star. Figure~\ref{fig:LCVR} shows the bolometric (left panel) and $g'$-band
(right panel) LCs for models with progenitor radii of 50, 100, 150 and
200 $R_\odot$. All of these configurations have the same He core taken
from the He4 model,  
and they were constructed in a similar way as 
He4R270, i.e.smoothly attaching 
an H-rich envelope to the core (see Figure~\ref{fig:rho}). We
denote these models as He4R50, He4R100, He4R150 and He4R200. As seen from the
figure, it is clear that models with $R \approx 200$ $R_\odot$ are
more consistent with the early-time data. This finding is not affected
by the systematic uncertainty in the luminosity that would arise from an error
of 1 Mpc in the distance.

We conclude this analysis by claiming that a
progenitor with radius similar to that of a YSG star, as suggested
from pre-SN detections, is compatible with the early observations of
SN~2011dh. Moreover, we find that radii much smaller
than 200 $R_\odot$ fail to reproduce the observations.

\section{Discussion}
\label{sec:discussion}
\subsection{Single versus binary progenitor}
  SNe~IIb require the hydrogen-rich envelope of the progenitor star to be
  almost completely removed before the explosion. Two
  alternative mechanisms of envelope removal have been proposed to
  explain the progenitors of SNe~IIb, Ib, and Ic, thereby called
  ``stripped-envelope SNe'': (1) strong stellar winds in massive
single stars, and (2) mass transfer in close binary systems.  
In the first scenario  a very massive star
with a main sequence mass $\gtrsim 30 \,M_\odot$  is required
for the mass-loss rate to be large enough
\citep{2003ApJ...591..288H,2009A&A...502..611G}. This type of star   
has a He core mass $\gtrsim 8 \, M_\odot$ previous to the
explosion. The upper limit of the 
  main-sequence mass may be even larger according to recent stellar wind
mass-loss rates 
\citep[see][]{2005A&A...438..301B,2006A&A...452..295E,2006ApJ...637.1025F}. In
the 
  binary scenario, less massive stars are allowed with He core masses
  prior to the explosion in the range of 3--6 $M_\odot$
  \citep{1993Natur.364..509P,2010ApJ...725..940Y}. In the previous
  section we showed that such He-core mass range is in very good
  agreement with the observations of SN~2011dh. 
 
To further test the possibility of a single-star progenitor, we
calculated a model based on a progenitor with main sequence mass
of 25 $M_\odot$ which forms a He core of 8 $M_\odot$ previous to the
explosion (we call this model He8). 
In Figures~\ref{fig:LC1} and \ref{fig:velo} we show the LC
and $v_{\mathrm{ph}}$, respectively, for model He8
using the same \Ni\, mass and distribution as found for the He4 model
of section~\ref{sec:SLC} but with a larger explosion energy of $E=2$ foe
in order to reproduce the peak luminosity. Clearly, this model does not agree
well with the observations. While decreasing the explosion energy can 
improve the match to the expansion velocities, it would worsen the fit
to the LC irrespective of the \Ni\ mixing assumed. Note that the
timing of the second peak imposes an important constraint on the He
core mass. More massive helium stars reach the LC maximum at later
times because the heat produced 
by radioactive decays takes longer to diffuse out. The He8 model is
too massive to produce the second maximum at $\approx 20$ days as
observed for SN~2011dh, even assuming the most extreme \Ni\,
mixing. 

This situation cannot be remedied by assuming a different
distance to M51 within the uncertainty of 1 Mpc. This would move all
the LC points systematically upward or downward by $0.12$ dex, but it
would not change the shape of the LC, i.e. its width and peak timing.
Therefore our hydrodynamical modeling indicates that the He
core mass must be lower than 8 $M_\odot$, which favors a binary
origin for SN~2011dh.

Recently it has been suggested that the binary channel is
preferred to explain most of the stripped-envelope SNe.
This is supported for example by
population studies compared with the observed rates of SNe
\citep{2008MNRAS.384.1109E,2011MNRAS.412.1522S}.
In the case of SNe~IIb the binary picture is
further strengthened for two reasons:
Firstly, stellar evolution of single stars requires a
precise fine tuning of the initial parameters to leave a thin H
envelope previous to the explosion. This is much more naturally
explained in a binary context
\citep{1993Natur.364..509P,1994ApJ...429..300W,2010ApJ...725..940Y}. Secondly, 
 the detection of the binary companion in the case of the type IIb
 SN~1993J, and possibly also in the case of SN~2001ig. A K0 supergiant
 star was identified as 
the progenitor of SN~1993J \citep{1994AJ....107..662A}. However, the
pre-SN photometry showed an excess in the UV and B bands that was
associated with  a blue companion star. Approximately a decade after the
explosion the blue supergiant companion was confirmed
\citep{2004Natur.427..129M,2009Sci...324..486M}. In the case of
  SN~2001ig, a possible companion star was reported by 
\citet{2006MNRAS.369L..32R}. 

The observed pre-SN photometry of SN~2011dh fits  well
the spectral energy distribution (SED) of a single YSG star
\citep{2011ApJ...739L..37M,2011ApJ...741L..28V}. Therefore, for the
binary sceneario to apply, the companion flux should have no appreciable
effect on the pre-SN photometry. We have performed binary evolution 
calculations to test the possibility of systems that
are compatible with the pre-SN observations. For this purpose,
we employed the interacting-binary code
developed by \citet{2003MNRAS.342...50B}. The code was initially
constructed for low-mass systems and recently extended
to compute the evolution of massive stars by  including nuclear
reactions until the end of oxygen burning. Further details of these  
calculations are presented in our companion paper \citet{BB12}. 

\begin{figure}
\begin{center}
\includegraphics[width=6cm]{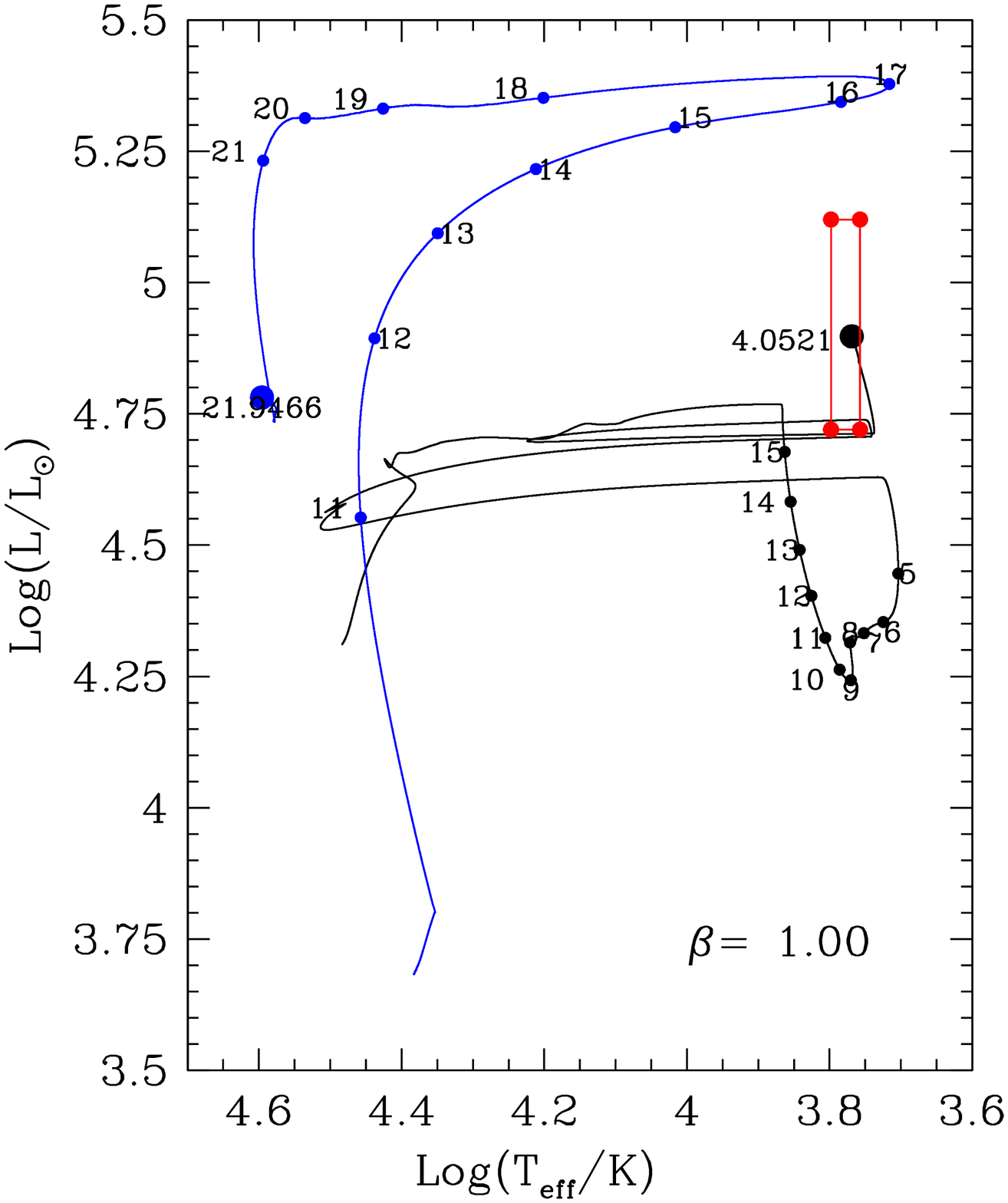}\includegraphics[width=6.cm]{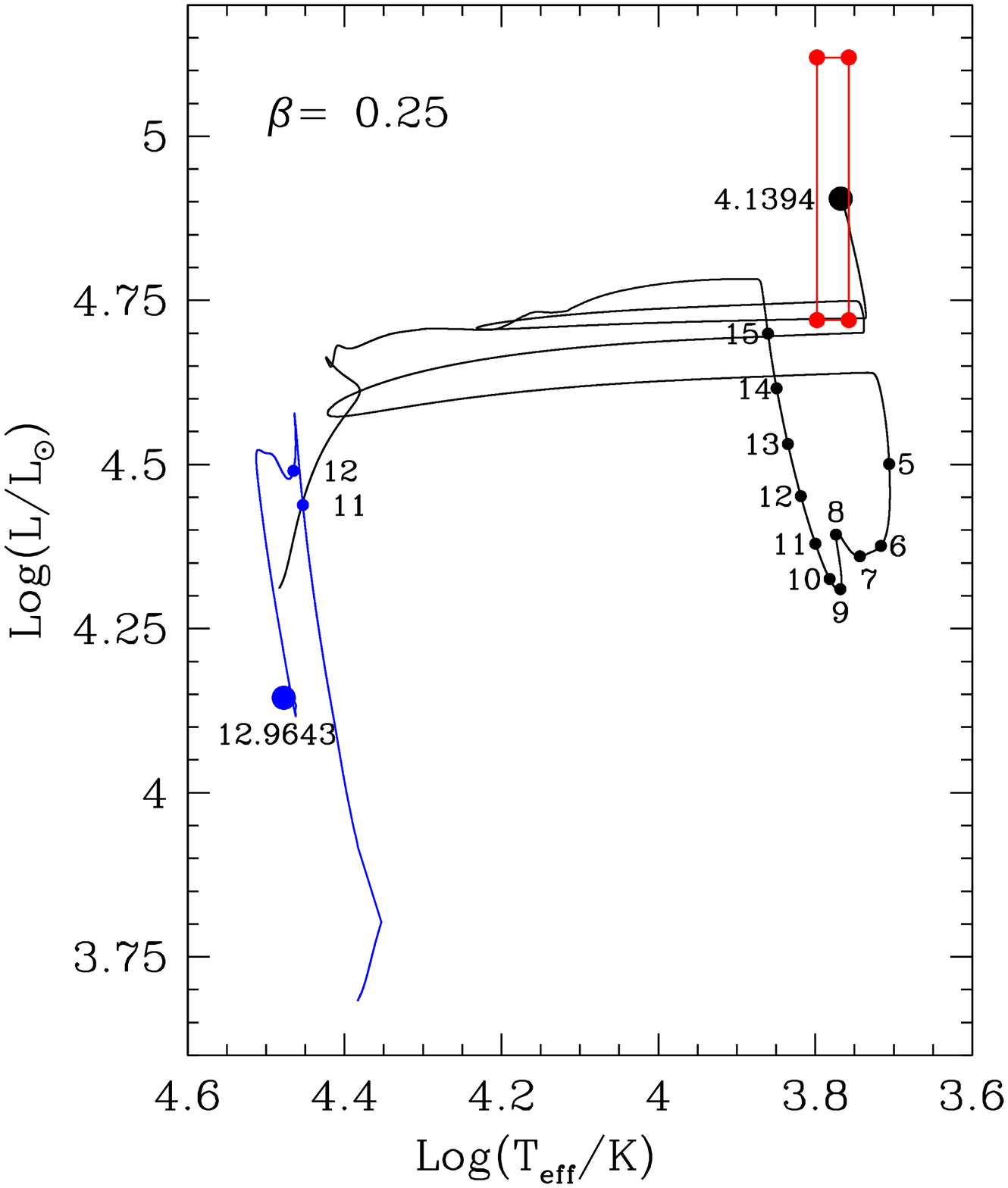} 
\caption{Evolutionary tracks in the Hertzsprung-Russell diagram for
  both components of a binary system with solar composition. The stars
  have $M_{\mathrm{ZAMS}}$ of 16 $M_\odot$ and 10 $M_\odot$ and an
  initial period of 150 days. {\bf Left panel:} assuming conservative
  mass transfer ($\beta=1$). {\bf Right panel:} non-conservative mass
  transfer ($\beta= 0.25$). Labeled dots along the tracks indicate the masses
  of the stars (in solar masses) while mass-transfer by Roche-Lobe
  overflow occurs. The primary star (black line) ends its evolution
  with a mass $\approx 4 \, M_\odot$ and with effective temperature and
  luminosity consistent with the YSG star detected in pre-SN images
  (red rectangle), independently of the value of $\beta$. However, the
  evolution of the secondary star (blue line) strongly depends on 
  the assumed accretion
  efficiency. In the conservative case, the bolometric luminosity of
  both stars at the moment of the explosion of the primary star (big
  dot) is similar, while for $\beta= 0.25$ the secondary is $\approx$ 2
  mag weaker than the primary at the end point. \label{fig:hrd}}
\end{center}
\end{figure}

\begin{figure}
\begin{center}
\includegraphics[scale=.60,angle=-90]{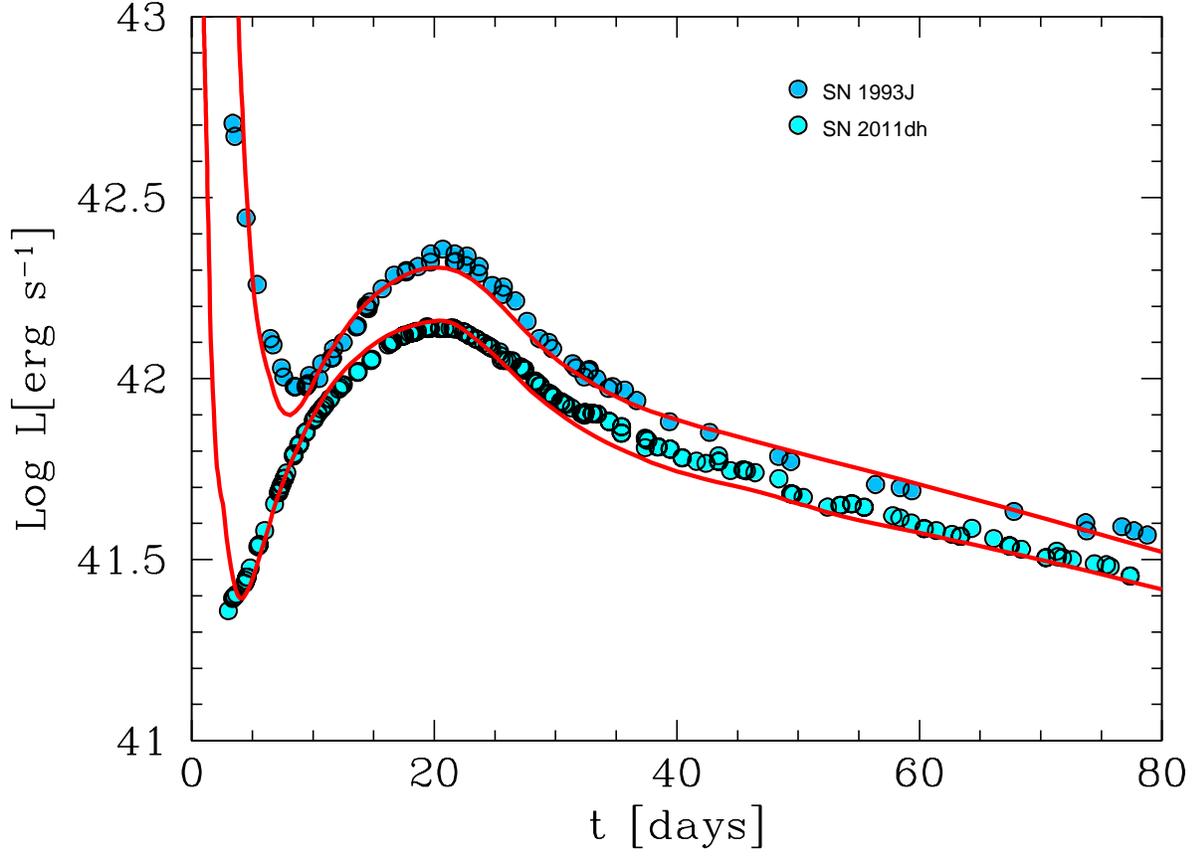}
\caption{Comparison between theoretical bolometric LCs (solid lines) and
observations (dots) for SN~1993J \citep{1994AJ....107.1022R} and
SN~2011dh \citep{E12}. In both cases a progenitor  was used with He core
mass of 4 $M_\odot$. Envelopes of different radii were attached to the
core to make $L$ and $T_{\mathrm{eff}}$ consistent with pre-explosion imaging of each
SN. The differences during the cooling phase are well explained as
differences in the size of the progenitor while the differences in the
main peak and tail are related with different different values of the
explosion energy and \Ni\, mass. \label{fig:LC93J}}

\end{center}
\end{figure}

Figure~\ref{fig:hrd} shows the evolutionary tracks in the 
H-R diagram for both components of a binary
system composed of a primary (donor) star with an initial mass of 16
$M_\odot$, a secondary (accreting) star of 10 $M_\odot$, and an
initial orbital period of 150 days. The conservative mass-transfer case
(\footnote{$\beta$ represents the fraction of material lost
by the donor star that is accreted by the secondary.}$\beta=1$)  is
shown in the left panel of the figure, where all the matter lost by the
primary star is accreted onto the secondary. 
The right panel shows a non-conservative case with an accretion
efficiency of 25\% ($\beta=0.25$). With this configuration, 
independently of the adopted value of $\beta$, the primary star ends
its evolution within the region of the H-R diagram defined by the
pre-SN photometry and its uncertainty. Also note that
the mass of the primary star prior to the explosion is $\approx 4
M_\odot$, which is consistent with the hydrodynamical modeling
presented in the previous section. 

Furthermore, in all the above calculations some hydrogen remains in
the envelope of the primary star at the moment of the explosion with a
similar mass of $\approx 4 \times 10^{-3} M_\odot$. This H mass is
high enough to be detected in the spectra and therefore it is
consistent with the SN~IIb classification \citep{2011MNRAS.414.2985D}. 

To study the effect of the secondary on the
pre-explosion photometry we used atmosphere models of
\citet{Kurucz93}\footnote{The Kurucz stellar atmospheres atlas was
  downloaded from
  \url{http://www.stsci.edu/hst/observatory/cdbs/k93models.html}} for
the corresponding $T_{\mathrm{eff}}$, surface gravity 
  and radius of both binary components to compute synthetic magnitudes
  through the observed HST bandpasses. As can be seen in
  Figure~\ref{fig:hrd}, the secondary star is significantly bluer than
the primary at the time of explosion. Thus, the largest contribution
to the flux of the system from the secondary is in the bluest band,
F336W. For both values of $\beta$ the contribution of the secondary in
the rest of the bands is $<$ 4\%. In the conservative case
($\beta=1$) the secondary would increase the flux of the system in the
F336W band by 50\%, producing a marginal $1.5$ $\sigma$ detection,
considering the measurement uncertainty in this band. In the
 non-conservative case with $\beta=0.25$ the 
 contribution of the secondary in the bluest band is only of 30\% of
 the total flux which falls well inside the photometric uncertainty
 ($0.6$ $\sigma$). 

If the binary scenario really applies to the case of SN~2011dh,
the companion star should be recovered in future observations,
 when the  SN becomes faint enough. From the synthetic photometry
  described above and assuming a distance to M51 of $7.1$ Mpc 
\citep{2006MNRAS.372.1735T} and no host-galaxy extinction, 
the expected brightness of the
  secondary in the F336W band is $24.9$ mag and $24.2$ mag 
for $\beta=0.25$ and $\beta=1$, respectively.
In the redder bands, the brightness lies within 26--27
mag. A search for the secondary star using the HST when the SN fades
enough is feasible and can serve to ultimately test the binary
scenario.

\subsection{Compact versus extended progenitor}
\label{sec:compact}

In the literature, a more compact progenitor has been
suggested for SN~2011dh than the YSG object detected at the SN
position
\citep{2011ApJ...742L..18A,2012ApJ...752...78S,2011ApJ...741L..28V}.  
This was based on (1) a  comparison between the early
decline rate in the LC of SN~1993J and SN~2011dh, (2) an
incompatibility between the temperature
derived from an early-time spectrum and that estimated from analytic
expressions for an extended object, and (3) large shock velocity
derived from radio observations.  

Regarding the first point, while it is true that SN~1993J showed a
slower decline during the cooling phase as compared with SN~2011dh,
this can be explained by the difference in size of the proposed
progenitors. SN~1993J was suggested to have a red
supergiant progenitor with a radius of $\approx 600 R_\odot$
 \citep[][ $\mathrm{log}\, L/L_\odot = 5.1$ and
$\mathrm{log}\, T_{\mathrm{eff}}= 3.63$]{2004Natur.427..129M}, whereas
the YSG star that was proposed as the progenitor
of SN~2011dh has a smaller radius of $\approx 200 R_\odot$.
To test this statement we calculated a model for SN~1993J based on
the He4 progenitor with an envelope attached to make $L$ and
$T_{\mathrm{eff}}$ consistent with the 
pre-explosion photometry of SN~1993J
\citep[see][]{2004Natur.427..129M}. Figure~\ref{fig:LC93J} shows a 
comparison between our model and the observations of SN~1993J
\citep{1994AJ....107.1022R} assuming the explosion date was 1993 March
$27.5$ UT. An explosion energy of $E= 1.2$ foe and a \Ni\, mass of $0.084$ 
$M_\odot$ were used in this simulation. Such values are consistent
with those adopted by \citet{1994ApJ...420..341S}. We also show the 
He4R270 model and the 
observations of SN~2011dh for a clear comparison between both SNe
and models. From the figure it is clear that the difference in
radius is enough to explain the difference in decline rate of the
early light curves between both SNe.

With respect to the temperature discrepancy of the second point above,
using hydrodynamical models in \S~\ref{sec:ELC} we did
not find any incompatibility with the spectrum temperature at $t\approx
2.8$ days when assuming an extended object. Note that our models show 
that the differences in temperature for $t\gtrsim 2$ days between compact and  
extended progenitors are too small to discriminate between these scenarios.
In our modeling we are able to determine the radius and thus set a
lower limit of about 200 $R_\odot$ (see Section~\ref{sec:ELC}) based on
the comparison with the early LC and not with the temperature.

Regarding the issue of the radio properties, the shockwave velocity
appears to be too high for an extended object based on
the picture presented by \citet{2010ApJ...711L..40C}. However, the
connection between the shock velocity and the  
physical radius of the progenitor is not direct and relies on several
assumptions (see \citet{2012ApJ...752...78S};
\citet{2012arXiv1201.0770K}). Therefore, we think that   
this argument alone  cannot be used to rule out the extended
progenitor scenario. 
Moreover, it is worth noting that the mass-loss rate estimated from
radio observations by \citet{2012arXiv1201.0770K} is 
compatible with a YSG progenitor and that no 
 variability in the radio light curve, as  has been observed for
   some compact SNe~IIb
   \citep{2004MNRAS.349.1093R,2006ApJ...651.1005S}, was reported to
 date.

\section{Conclusions}
\label{sec:conclusion}
We have calculated a set of hydrodynamical models applied to
stellar-evolution progenitors in order to study the nature of SN~2011dh. 
Comparing our models with the observed bolometric LC during the second peak 
and with line expansion velocities we found that a progenitor with He core mass 
of  $3.3$--4 $M_\odot$, an explosion
energy of  $6-10\times 10^{50}$ erg, and 
a \Ni\, mass of $\approx$ $0.06$ $M_\odot$ reproduce very well the 
observations assuming a distance of $7.1$ Mpc to M51. This type of
model is consistent with a main sequence mass  between 12 and 15
  $M_\odot$.  Remarkably, the range of mass found with our hydrodynamical
  modeling is in very good agreement with the estimates from two other
  independent methods, i.e. pre-SN imaging and
  stellar population analysis. This is different from the situation
  generally encountered for SNe~IIP where LC modeling 
  estimates main sequence masses that are higher up to a factor
  of two than those estimated from pre-SN imaging
  \citep{2008A&A...491..507U,2009ARA&A..47...63S}.

 We have studied the effect of the progenitor radius on the early LC 
and temperature evolution. We found that a progenitor with radius
 similar to that of the YSG star detected in the pre-SN images 
is compatible with the early observations of SN~2011dh without
contradiction with the temperature that is derived from the spectrum,
as opposed to what \citet{2011ApJ...742L..18A} found using
analytic models. Furthermore, progenitors
with radii $< 200 \, R_\odot$ fail to 
reproduce the early $g'$-band LC. Although our hydrodynamical models
show differences in the temperature evolution between extended and
compact progenitors, these differences are less marked than those
predicted by analytic expressions and they are almost unnoticeable for 
$t\gtrsim$ 2 days. Therefore, the spectrum temperature at $t\approx
2.8$ days is not useful in this case to discriminate between compact
and extended progenitors.

We have tested {\em and ruled out} progenitors with He core
 masses   $\gtrsim 8 M_\odot$ which correspond to 
$M_{\mathrm{ZAMS}} \gtrsim 25 \, M_\odot$. Considering the limitations at such
 stellar masses of single-star winds to expel
 the H-rich envelope almost entirely, as required for SNe~IIb, this
 result is in favor of a binary origin for SN~2011dh.

We have also performed binary evolution calculations with mass
  transfer to test the 
possibility of systems that are compatible with the pre-SN
observations of SN~2011dh.  We have shown that a system with 
16 $M_\odot$ + 10  $M_\odot$ and an initial period of 150 days
predicts that the primary star ends its evolution in the H-R diagram
at the right position as compared with the YSG star detected in the
pre-SN images \citep{2011ApJ...739L..37M,2011ApJ...741L..28V}. Furthermore, 
the He mass of the primary at the end of the evolution was $\approx 4 M_\odot$,
which is consistent with our hydrodynamical modeling. The binary
evolution calculations further 
predict that some hydrogen mass of $\approx 4 \times 10^{-3} M_\odot$ is left  
in the envelope of the primary star, which is required to produce a
SN~IIb. 

To test the binary scenario we have studied the effect of the
putative companion star on the pre-explosion photometry in
comparison with the observations. Because the secondary star is
predicted to be much hotter than the primary star, we found that the
largest effect appears in the blue and UV filters. The contribution of
the secondary to the flux in the F336W band, however,
is marginal, at the $1.5$ $\sigma$ level.
The contribution is further decreased to the $0.6$ $\sigma$ level when
non-conservative mass accretion is considered. However, our
  predictions can be tested in a few years time by a search for a blue
object at the location of the SN. 

\acknowledgments
This research has been supported in part by the Grant-in-Aid for
Scientific Research of MEXT (22012003 and 23105705) and 
JSPS (23540262) and by World Premier International Research Center
Initiative, MEXT, Japan. SB is partially supported by the PRIN-INAF
2009 with the project "Supernovae Variety and Nucleosynthesis Yields".

\clearpage

\clearpage

\begin{deluxetable}{lccccccc}
\tablecaption{Physical parameters of the explosion models\label{tbl-1}}
\tablewidth{0pt}
\tablehead{
\colhead{}  & \colhead{$M_\mathrm{ms}$\tablenotemark{a} } & \colhead{$M_\mathrm{He}$} & \colhead{$R_\star$}&
\colhead{$M_\mathrm{cut}$}& \colhead{$M_\mathrm{ej}$}& \colhead{$E$} &
 \colhead{\Ni\,mass} \\
\colhead{Model} & \colhead{[$M_\odot$] } & \colhead{[$M_\odot$]} & \colhead{[$R_\odot$]} &
\colhead{[$M_\odot$]} & \colhead{[$M_\odot$]} & \colhead{[foe]} & \colhead{[$M_\odot$]} 
}  
\startdata 
He3.3    &  12   &  3.3  & 2.5 & 1.5  & 1.8 & 0.6  & 0.065   \\
He4      &  15   &  4.   & 2.4 & 1.5  & 2.5 & 1.   & 0.06     \\
He5      &  18   &  5.   & 1.8 & 1.6  & 3.4 & 1.2  & 0.065   \\
He8      &  25   &  8.   & 1.3 & 1.8  & 6.2 & 2.   & 0.065   \\    
\enddata
\tablenotetext{a}{Main sequence mass} 
\end{deluxetable}

\end{document}